\documentclass[twocolumn]{aa}  
\usepackage{natbib}
\usepackage{twoopt}

\bibpunct{(}{)}{;}{a}{}{,} 
 
\makeatletter
\newcommandtwoopt{\citeads}[3][][]{\href{http://adsabs.harvard.edu/abs/#3}%
{\def\hyper@linkstart##1##2{}%
\let\hyper@linkend\@empty\citealp[#1][#2]{#3}}}
\newcommandtwoopt{\citepads}[3][][]{\href{http://adsabs.harvard.edu/abs/#3}%
{\def\hyper@linkstart##1##2{}%
\let\hyper@linkend\@empty\citep[#1][#2]{#3}}}
\newcommandtwoopt{\citetads}[3][][]{\href{http://adsabs.harvard.edu/abs/#3}%
{\def\hyper@linkstart##1##2{}%
\let\hyper@linkend\@empty\citet[#1][#2]{#3}}}
\newcommandtwoopt{\citeyearads}[3][][]%
{\href{http://adsabs.harvard.edu/abs/#3}
{\def\hyper@linkstart##1##2{}%
\let\hyper@linkend\@empty\citeyear[#1][#2]{#3}}}
\makeatother

\usepackage{graphicx}
\usepackage{color}
\usepackage[varg]{txfonts}
\usepackage{hyperref}   

\usepackage{arydshln}
\usepackage{multirow}

\usepackage{float} 

\begin{document} 
   \title{VarIabiLity seLection of AstrophysIcal sources iN PTF (VILLAIN) \\
   I. Structure function fits to 71 million objects 
}

   \author{
           S. H. Bruun\inst{1}
           \and A. Agnello\inst{1,2}
           \and
           J. Hjorth\inst{1}
           }

   \institute{DARK, Niels Bohr Institute, University of Copenhagen,
              Jagtvej 128, 2200 Copenhagen N, Denmark
              \and
              STFC Hartree Centre, Sci-Tech Daresbury, Keckwick Lane, Daresbury, Warrington (UK) WA4 4AD
              \\
              \email{ sofie.bruun@nbi.ku.dk, adriano.agnello@nbi.ku.dk, jens@nbi.ku.dk}
              }

    \date{Accepted by A\&A April 11 2023}

    \titlerunning{Variability selection in PTF: I. Processing of lightcurves}
    
  \abstract
   {Lightcurve variability is well-suited for characterising objects in surveys with high cadence and long baseline. This is especially relevant in view of the large datasets to be produced by the Vera C. Rubin Observatory Legacy Survey of Space and Time (LSST). 
   } 
   {We aim to determine variability parameters for objects in the Palomar Transient Factory (PTF) and explore differences between quasars (QSOs), stars and galaxies. We will relate variability and colour information in preparation for future surveys.}
   {We fit joint likelihoods to structure functions (SFs) of 71 million PTF lightcurves with a Markov Chain Monte Carlo method. For each object, we assume a power law SF and extract two parameters: the amplitude on timescales of one year, $A$, and a power law index, $\gamma$. With these parameters and colours in the optical (Pan-STARRS1) and mid infrared (WISE), we identify regions of parameter space dominated by different types of spectroscopically confirmed objects from SDSS.
   Candidate QSOs, stars and galaxies are selected to show their parameter distributions.}
   {QSOs have high amplitude variations in the $R$ band, and the strongest timescale dependence of variability. Galaxies have a broader range of amplitudes and low timescale dependency. 
   With variability and colours, we achieve a photometric selection purity of 99.3~\% for QSOs. 
   Even though hard cuts in monochromatic variability alone are not as effective as seven-band magnitude cuts, variability is useful in characterising object sub-classes. 
   Through variability, we also find QSOs that were erroneously classified as stars in the SDSS.  
   We discuss perspectives and computational solutions in view of the upcoming LSST. }
   {}

   \keywords{Methods: statistical -- quasars: general -- surveys -- Techniques: photometric -- Methods: data analysis -- Galaxies: active}

   \maketitle
   
\section{Introduction}
Large, wide-field surveys 
allow us to identify rare objects, study parameter distributions of different object classes and select sources for further examination. Spectroscopy can produce high quality classifications, but often only photometry is available. With the prospect of deep, 10-year lightcurves from the Vera C. Rubin Observatory Legacy Survey of Space and Time \citep[LSST;][]{LSST}, it is important to explore photometric selection based on lightcurve variability as a function of timescale for a large dataset. 

This method is particularly suitable for analysing certain types of objects with distinctive variability parameters, e.g. for distinguishing quasars (QSO) from stars \citep{Bergh,schmidt,Ulrich,SDSS_PTFQSOselection}. 
Variability can also advance understanding of the physical nature of the objects, as the mechanisms behind variability operate on different timescales \citep{schmidt_col} -- such as QSO accretion disk instabilities \citep{Rees1984} or changes in accretion rate \citep{Hopkins2006}. 

With identification of distant QSOs, we can study structure formation in the early universe  
and the cosmological parameters determining its expansion, e.g. through measurements of baryon acoustic oscillations (BAO) in the Ly$\alpha$ forest \citep{alam,bao_lya,turner,song,QSO_cosmoprinc}. \citet{SDSS_PTFQSOselection} selected a fraction of the SDSS eBOSS QSO targets for BAO based on variability in lightcurves from the Palomar Transient Factory \citep[PTF;][]{PTF,Rau}. QSOs are also useful for redshift-drift tests of cosmic acceleration \citep{redshiftdrift,kim,alves,Loeb1998}. 
Especially lensed QSOs are interesting because they can be used for time-delay cosmography, which works better with highly variable sources \citep{Refsdal,treu}. 
Lensed QSOs can be discovered as extended objects with high variability \citep{kochanek}.

Faint photometric standards are also of special interest. Large future observatories such as the 
Vera C. Rubin Observatory, the Extremely Large Telescope \citep{ELT}, the Thirty Meter Telescope \citep{TMT} and the Giant Magellan Telescope \citep{GMT} will observe to higher magnitudes than the typical magnitudes of standard stars. These have magnitudes of $11.5<V<16$ in the \citet{Landolt} sample, which has, 
along with the Stetson database of secondary standards \citep{stetson2000,stetson2019}, since been expanded and curated as described in \citet{Gaia_stdstar}. The combined sample mainly includes sources with $13<V<21$. 
The distribution on the sky is not uniform, however, and knowing variable from non-variable sources is useful for calibration even at lower magnitudes, such as the 20.6 R-band limit of the PTF. 

There are several ways of characterising variability,  and they can be even more useful than colours for selection of AGN \citep{cicco2021}. \citet{lowmassAGN} demonstrated the potential in PTF variability for selection of AGN in low mass galaxies by fitting 50,000 lightcurves with a damped random walk (DRW) model. \citet{ZTFAGN} found variable AGN in the Zwicky Transient Facility \citep[ZTF;][]{ZTF} using a combination of variability measures. 

In this paper, we apply structure functions (SF) to explore the variability of objects. This is a well-known technique, first applied in astronomy by \citet{Simonetti}, which is computationally efficient for large sets of data with gaps \citep{SFgaps}. SF models describe the difference in magnitude $\Delta m$ across a time interval $\Delta t$. We specifically consider a power law model, as done by e.g. \citet{hook1994}, which is often applied for AGN and QSOs. Both SF and DRW models 
pose challenges -- even 20 year baselines cannot constrain DRW models completely \citep{suberlak, stone}.

\citet{Sanchez17} found the Bayesian SF defined by \citet{schmidt} to be the best and most stable one for noisy lightcurves with irregular sampling. \citet{schmidt} applied power-law SFs to lightcurves from The Sloan Digital Sky Survey (SDSS) and defined a variability parameter region for selection of QSO candidates with criteria chosen by eye. This gave relatively complete and pure candidate sets for SDSS S82 lightcurves, but with worse performance for lightcurves from Pan-STARRS1 \citep[PS1;][]{PS1survey}. 
The performance was achieved 
on sets of known QSOs, F/G stars and RR Lyrae, with stars outnumbering QSOs by a factor of 30. 

In this paper, we aim to analyse the results and performance of a similar method but applied to the entire PTF survey. 
We will show where variability is most effective in breaking degeneracies and which forms of variability are common among different types of objects. Specifically,
matching with spectroscopic SDSS classifications of QSOs, stars and galaxies, which we assume as ground truth, we can evaluate new variability selection criteria and compare with those of \citet{schmidt}\footnote{This analysis expands the first queries and parameter exploration presented by \citet{Msc}, MSc thesis (unpublished).}. By including as many objects from the PTF survey as possible, the properties of the total lightcurve sample are more representative of PTF sources than if we only included e.g. specific stellar subtypes.

Another approach to photometric selections is based on colours.
We will examine colour-colour and colour-magnitude diagrams of a spectroscopically confirmed sample by matching magnitudes in mid-IR from the Wide-field Infrared Survey Explorer \citep[WISE;][]{WISE} 
and in the optical from PS1. This will allow us to examine simple selection criteria based on these. 

We describe the datasets (and the data cleaning process) used in this work in Sect. \ref{sec:data}.
Sect. \ref{sec:methodology} introduces the methodology. In Sect. \ref{sec:sf}, we define models of the PTF lightcurve SFs, and we fit them in Sect. \ref{sec:fitting}. The metrics used for evaluation of selection of objects are defined in in Sect. \ref{sec:evaluation}. The results of the SF fitting are plotted and described in Sect. \ref{sec:results}. Based on the variability parameters and matched colour information, we choose photometric selection criteria and examine the properties of selected objects in Sect. \ref{sec:photoclass}. We discuss them in Sect. \ref{sec:discussion} and summarise the findings in Sect. \ref{sec:conclusion}. 

\section{Data} \label{sec:data}

\subsection{PTF} \label{sec:data:PTF}

The Palomar Transient Factory was a project with the Palomar 48 Schmidt telescope at Palomar Observatory in California. This includes imaging in DR1, DR2 and DR3 \footnote{as the Intermediate Palomar Transient Factory for DR3} from 1 March 2009 to 28 January 2015 covering 20\ 000 deg$^{2}$ \citep{PTF,Rau}\footnote{www.ptf.caltech.edu/iptf}. 
The PTF lightcurve database is based on a subset of these images with data from 598\,975\,024 objects in $R$ and $g$. 

Since most data points are in $R$, these are chosen for the analysis of this paper (as \texttt{mag\_autocorr} and \texttt{magerr\_auto}). The $R$ band is adapted from the 658 nm Mould-$R$ filter described in \citet{MouldR}, has a limiting magnitude of 20.6 and photometry is in the AB magnitude system \citep{PTF}.
From this database, we also consider the timestamps, PTF object ID, RA, Dec and median $R$ -- queried in 648 patches of $10$\degr$\times10$\degr\ via the IRSA Catalog Search Tool\footnote{PTF team, DOI: 10.26131/IRSA156, url: https://irsa.ipac.caltech.edu/Missions/ptf.html,  IPAC}. 310 large patch files are split into 6340 files, keeping data points with the same object ID together. File by file, we group data points by ID to assemble lightcurves. These are cleaned and fitted in Sects. \ref{sec:cleaning} and \ref{sec:sf} in parallel using 14 computing nodes for six months. 

\subsection{SDSS} \label{sec:data:SDSS} 
The Sloan Digital Sky Survey DR17 \citep{SDSSIV_overview,SDSSDR17} 
includes 5.789 million spectra \citep{SDSSspec1,SDSSspec2,SDSSspec3} 
from the Apache Point Observatory in New Mexico, USA \citep{SDSS_telescope}. 866\,338 QSOs, 962\,162 stars and 2\,790\,253 galaxies are spectroscopically confirmed based on DR17 from February 2021 (with data both in PhotoObj and SpecObj). 
Their coordinates, redshifts and spectroscopic classes are queried via CasJobs on SciServer \citep{SciServer}. 
We cross-match by creating a $k$-d tree \citep{kdtree} of the PTF coordinates. 
Then we query the tree to find the nearest SDSS object within two arcseconds, if it exists. 
By inspecting SDSS images\footnote{using the Image List tool at skyserver.sdss.org/dr17/VisualTools/list}, we find that this radius best avoids spurious matches.  

\subsection{WISE} \label{sec:data:WISE}

The Wide-field Infrared Survey Explorer is a space telescope observing in the mid infrared \citep{WISE}. It had a cryogenic phase from Dec. 14, 2009 to Feb. 2011 and a post-cryogenic one since 2013 (NEOWISE). 
With the latest update from February 2021, the combined AllWISE program \citep{AllWISE} data release II/328 contains 748 million objects. 
 
We query WISE objects with the CDS X-Match Service \citep{Xmatch} within five arcseconds of the mean PTF coordinates to each object. From WISE \citep[‘vizier:II/328/allwise’;][]{AllWISE,vizier}, we get $W1$ (3.4 µm), $W2$ (4.6 µm), their errors and coordinates.
$W3$ and $W4$ are not included, as they are not deep enough to include most of the PTF sources.  
WISE magnitudes are in the Vega magnitude system \citep{WISE}. 
For each PTF object, only the closest WISE source is selected via a $k$-d tree like in Sect. \ref{sec:data:SDSS}. If this source contains multiple data points, we choose
the lowest magnitude, as that is the brightest detection, and save
the mean RA and Dec.

\subsection{PS1} \label{sec:data:PS1}

The Panoramic Survey Telescope and Rapid Response System includes two telescopes in Hawaii; Pan-STARRS1 and Pan-STARRS2. We use the $g$ (481 nm), $r$ (617 nm)  
and $z$ (866 nm) 
bands \citep{PS1_photometric} from the Pan-STARRS1 survey DR1 with 1.92 billion objects \citep[‘vizier:II/349/ps1’;][]{PS1survey,vizier}. This covers the sky at J2000.0 declination > $-$30 \degr, including the PTF footprint. The PS1 magnitudes are in the AB system.
Like for WISE, in PS1 we query within five arcseconds of PTF objects and select one set of magnitudes, magnitude errors and coordinates per source.  

\subsection{Data cleaning} \label{sec:cleaning}

Data cleaning is needed for reliable fitting results. For PTF, extreme magnitudes outside the range of $12<R<22$ are removed. 
If a lightcurve contains too few data points, fit parameters are difficult to constrain, and if the time span is too short, we cannot detect variability over long timescales. Therefore, we require that each lightcurve must contain at least 20 data points and span at least one year. At 80\degr < RA < 90\degr\ and 0\degr < Dec < +10\degr, lightcurves are so well sampled that to speed up the analysis in Sect. \ref{sec:fitting}, we select a maximum of 1500 random data points for some sources. 

Magnitude outliers affect variability fits, so we want to remove them, while preserving the real variability. 
A moving weighted median (MWM)  
is computed for each data point $i$ in each lightcurve. This is based on the magnitudes $R_{close,i}$ of the seven closest neighbouring data points in time within a window of five days (2.5 days to each side). 
Data points more than three standard deviations away from the MWM are removed, taking into account both the error of the data point and of the weighted median. A data point with magnitude $R$ and error $\sigma_R$ is removed if
\begin{align}
    \frac{|R-\text{MWM}|}{\sqrt{\sigma_R^2 + \text{MAD}^2}}&>3, 
\end{align}
where MAD is the Median Absolute Deviation. We compute the MAD of every part of the MWM as the weighted median of the absolute distances of the data points used in the computation of the MWM. That is, the MAD of a specific value of $\text{MWM}$  is based on the up to seven data points within the time window of data point $i$:
\begin{align}
    \text{MAD} = \text{weighted median}\left(|R_{close,i} - \text{MWM}|\right).
\end{align}
If fewer than seven data points are close enough to $i$ in time, they are still used, and the data point $i$ itself has a relatively higher weight compared to those few data points, leading to a higher probability of acceptance. This is desirable, as we have less information to base rejection on.

After data cleaning, the remaining lightcurves are distributed on the sky according to Fig. \ref{fig:Nobj_per_patch}, mostly covering the northern hemisphere and avoiding low Galactic latitudes.
Table \ref{tab:samples} contains sample statistics including objects matched in SDSS, WISE and PS1. 70\,920\,904 objects are analysed, spanning $365-2147$ days and containing $20-5579$ data points.

\begin{table}
      \centering
      \caption{Sample sizes.}
      \begin{tabular}{l l }
          \hline\hline
          Sample & Counts \\
          \hline
          Full PTF lightcurve database & 598\,975\,024 \\
          Cleaned PTF R-band sample & 70\,920\,904 \\
          Matched & \\
          \hspace{0.5cm}  SDSS spectra & 1\,748\,047 \\
          \hspace{0.5cm}  WISE        & 57\,007\,069 \\
          \hspace{0.5cm}  PS1         & 70\,891\,378 \\
          \hspace{0.5cm}  WISE \& PS1 & 56\,991\,591 \\
          \hspace{0.5cm}  All surveys & 1\,613\,916  \\
      \end{tabular}
      \label{tab:samples}
      \tablefoot{Sample sizes before and after selecting and cleaning PTF lightcurves and matching with sources in SDSS, WISE and PS1.}
\end{table}{}

\section{Methodology} \label{sec:methodology}
We describe the variability of each object by defining power law models of their structure functions. The models are fit to structure functions of individual objects to extract variability descriptors as fit parameters. These will be used for exploring relations in the data and photometric selection of object classes in Sects. \ref{sec:results}--\ref{sec:photoclass}; but first, we define metrics for evaluation of selection quality in Sect. \ref{sec:evaluation}.

\subsection{Structure functions} \label{sec:sf}

For each object, we analyse variability by comparing every pair of data points in its lightcurve. For each pair of data points $i$ and $j$, we have a difference in magnitude $\Delta m_{ij}$ and in time $\Delta t_{ij}$. By comparing all pairs, we find the timescale dependence of differences in magnitude with a SF \citep{Simonetti}, and model this variability.

The total effective variability $V_{\mathrm{eff}}$ of each object consists of intrinsic variability, which we describe with a model $V_{\mathrm{mod}}$, and noise, $\sigma_m$. Assuming the model describes the data well, $V_{\mathrm{eff},ij}$ is close to the observed $\Delta m_{ij}$. For $V_{mod}$, we choose a power law, so
\begin{align}
    V_{\mathrm{eff},ij}^2 &= V_{\mathrm{mod},ij}^2 + \sigma_{m,i}^2 + \sigma_{m,j}^2 \approx (\Delta m_{ij})^2, \\
    V_{\mathrm{mod},ij} &= A \left(\frac{|\Delta t_{ij}|}{t_0}\right)^\gamma,\ A\geq0. \label{eq:Vmod}
\end{align}
Eq. \ref{eq:Vmod} has two free parameters -- $A$ and $\gamma$. $A$ quantifies the amplitude, in units of magnitudes, of the variations on the timescale $t_0$, and the power law index $\gamma$ describes how the amplitudes depend on timescale. We choose $t_0 = 1 $ year (observed frame). One could correct for the factor of $1+z$ to find rest frame $\Delta t_{ij}$, but redshift information is limited, and SDSS redshifts are not independent of the spectroscopic labels we use for comparison.

The SF is simply the square of the model variability, following the notation of \citet{schmidt}, 
\begin{align}
    \text{SF} = A^2 \left(\frac{|\Delta t_{ij}|}{t_0}\right)^{2\gamma}.
\end{align}
We expect most objects to have $0<\gamma<1$. $\gamma>0$ shows that variability increases with timescale, and for $\gamma>1$ this is accelerating, which we would mostly expect to see for short lightcurves with highly uncertain $\gamma$. This could be objects with an overall positive or negative trend. $\gamma<0$ means most variability is found at short timescales, e.g. in transient tails, but we do not expect to include these objects due to the minimum observed time span of one year.  
Only $A$ consistently shows correlation with physical black hole parameters in the literature, namely an anti-correlation with the rest-frame emission wavelength, $\lambda_{rest}$, and the Eddington ratio, $L/L_{\mathrm{Edd}}$ \citep{cicco2022}.

\subsection{Fitting} \label{sec:fitting}

SF models are fitted to data using the emcee python package \citep{emcee} for affine-invariant sampling with Markov Chain Monte Carlo \citep[MCMC;][]{Goodman2010}. 
As the walkers of the MCMC jump to different $A$ and $\gamma$ values, they evaluate the likelihood of observing the lightcurve given the variability parameters. 
This likelihood is
\begin{align}
    L_{ij} = \frac{1}{\sqrt{2\pi V_{\mathrm{eff},ij}^2}} \exp{\left( -\frac{(\Delta m_{ij})^2}{2V_{\mathrm{eff},ij}^2} \right)}
\end{align}
for each pair of data points, assuming a gaussian distribution of $\Delta m_{ij}$. 
For the total log posterior, LP, of the lightcurve, we use the same prior $\ln(p)$ as in \citet{schmidt}:
\begin{align}
    \ln(p) &= \ln\left(\frac{1}{1+\gamma^2}\right)+\ln\left(\frac{1}{A}\right),\\
    \text{LP} &=\sum_{i,j}\ln(L_{ij}) + \ln(p).
\end{align}
The MCMC runs with eight walkers for 500 steps where the first 200 are discarded as burn-in. We find these values to balance accuracy and speed.
From each fit, we retain the median values of $A$ and $\gamma$ and their 16th and 84th percentiles as 1$\sigma$ uncertainties. 

\subsection{Evaluation metrics} \label{sec:evaluation}

To study parameter distributions of different object types, we photometrically select QSOs, stars and galaxies. 
The quality of selection criteria is evaluated on purity and completeness, estimated using the spectroscopically confirmed subset.
We compute 
these with the number of true positives, TP, false negatives, FN, and false positives, FP:
\begin{align}
    \mathrm{Completeness} &= \frac{\mathrm{TP}}{\mathrm{TP+FN}}, 
    & \mathrm{Purity}       = \frac{\mathrm{TP}}{\mathrm{TP}+\mathrm{FP}}. \label{eq:performance}
\end{align}

\section{Variability and colour distributions} \label{sec:results}

We present plots and statistics for two datasets: all fitted objects and those with spectroscopic classifications.  In Sect. \ref{sec:photoclass}, we also assemble photometric selections. The datasets allow us to compare distributions of three classes: QSOs, stars and galaxies.
The plots include variability parameters ($A$ and $\gamma$) and colour diagrams of $W2$ vs. $W1-W2$ and $g-r$ vs. $z-W1$. We know stars, galaxies and QSOs to have good separation in optical vs. infrared colours \citep{maddox,dustin}. $W1-W2$ was also used by \citet{WISE}, and \citet{assef} combined $W2$ and $W1-W2$ for selecting AGN.

We query and fit the data in batches as described in Sect. \ref{sec:data:PTF}. For 6340 files with approximately a million lightcurve data points per file, the batch processing time is about an hour on a computing node with 32 cores. The maximum time is 11 days for one file.

\subsection{Full PTF sample}

In Fig. \ref{fig:AllData}, we show parameter distributions for the full PTF sample. The top panel is a 2D-histogram of $A$ and $\gamma$ values, illustrating the variability of the 70\,920\,904 fitted lightcurves. 
We see two large clusters, at $\log A\sim-0.8$ and $\log A\sim-5$ (base 10). 
Variations of just $10^{-5}$ magnitudes over timescales of one year indicate that the values of the latter cluster are spurious. 

The middle and bottom panels of Fig. \ref{fig:AllData} display a colour-colour and a colour-magnitude diagram. As these are based on parameters in WISE and PS1, they show distributions for the 80~\% (56\,991\,591) of the PTF objects with matches in both of these surveys (see Table \ref{tab:samples}). 93~\% of QSOs, 75~\% of stars and 98~\% of galaxies have these matches. 

\subsection{Spectroscopically confirmed sample} \label{sec:specsample}

\begin{figure}
	\centering
    \resizebox{\hsize}{!}{\includegraphics{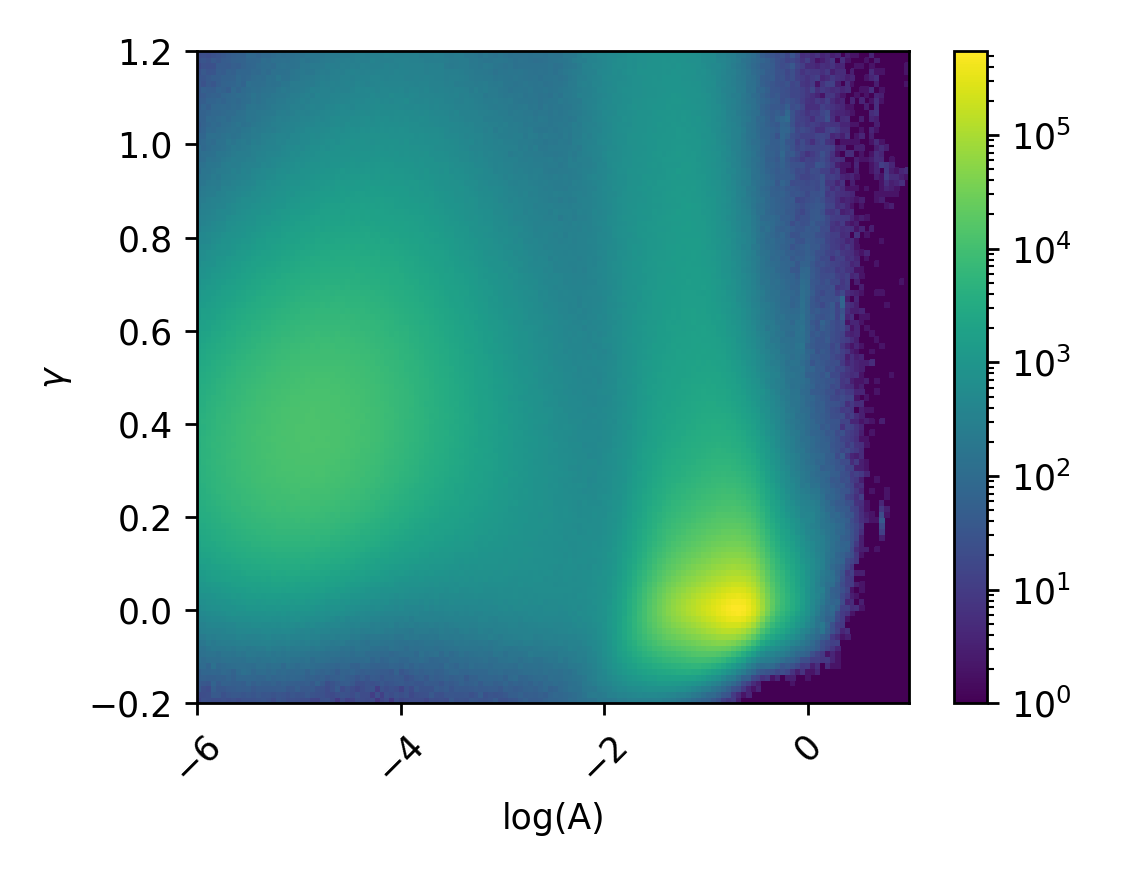}}
    \resizebox{\hsize}{!}{\includegraphics{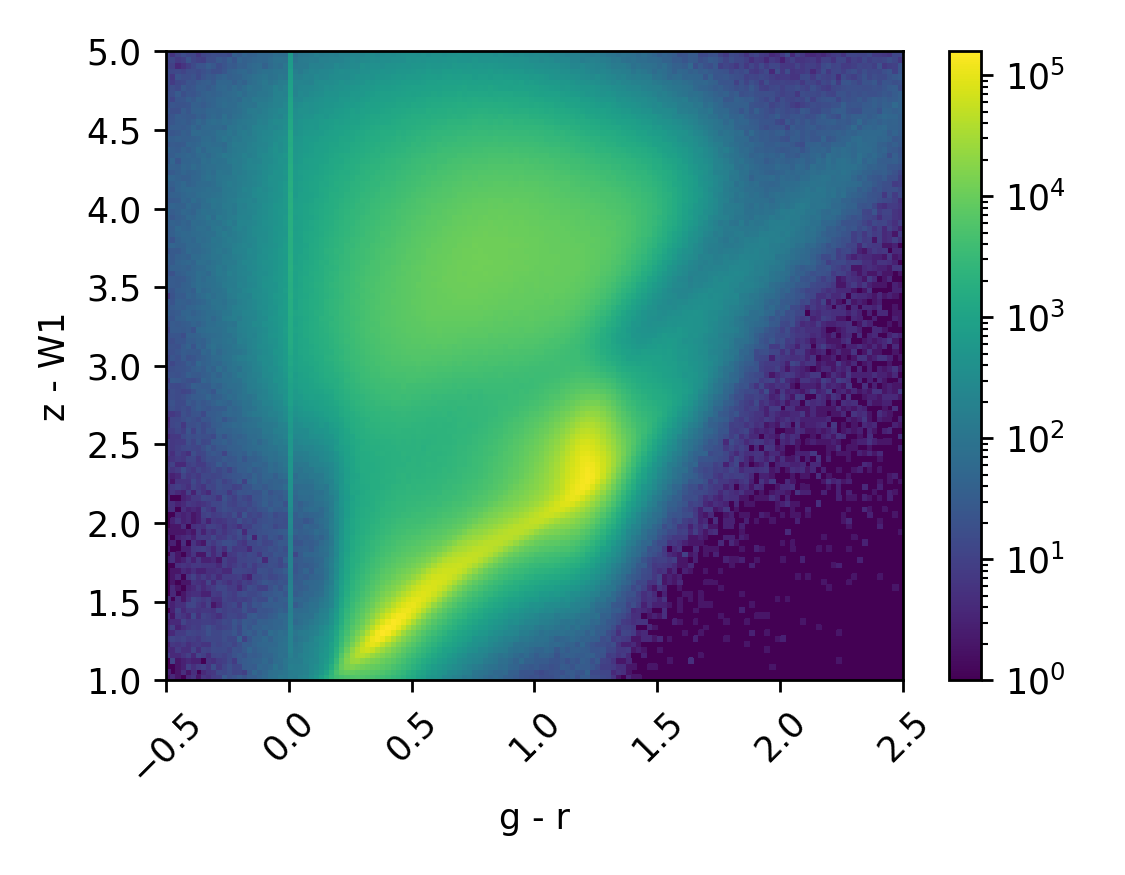}}
    \resizebox{\hsize}{!}{\includegraphics{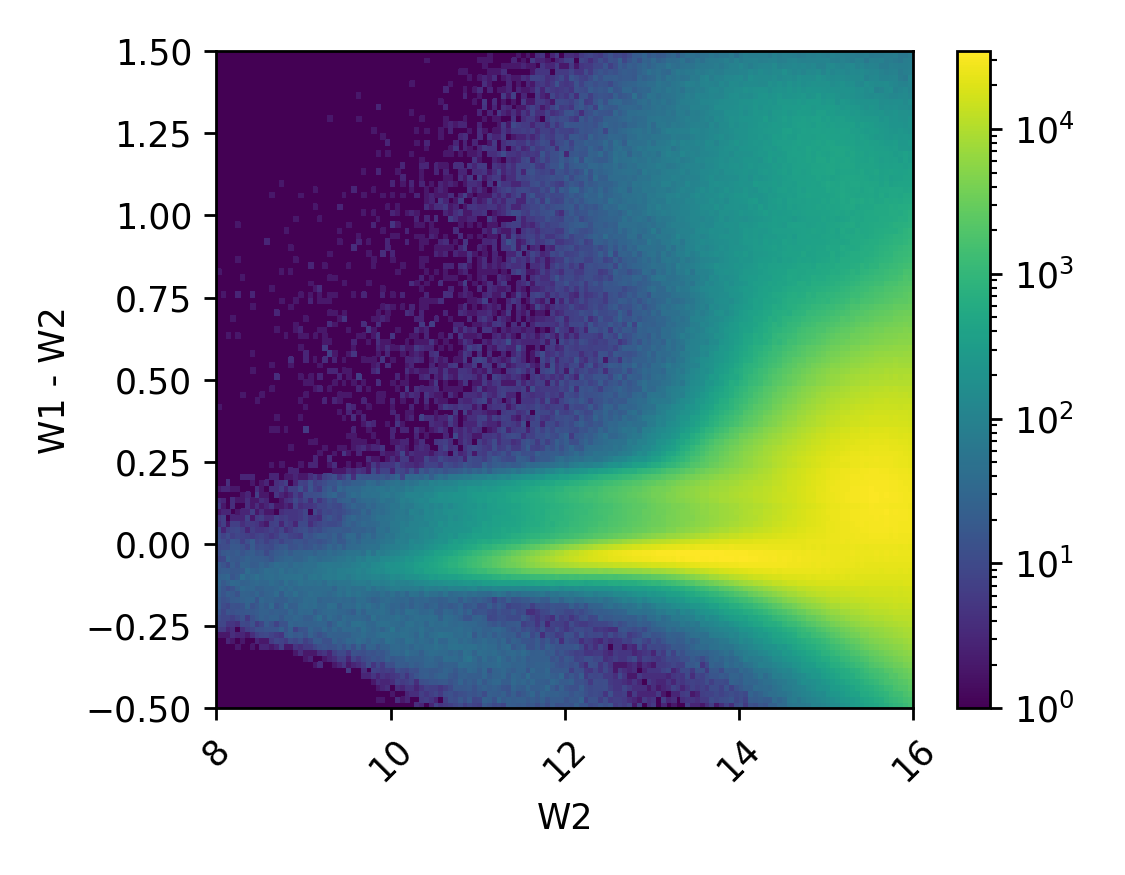}}
	\caption{Heatmaps of variability parameters (top),  $g-r$ vs. $z-W1$ (middle) and $W2$ vs. $W1-W2$ (bottom). These are the full parameter distributions of objects with lightcurves in PTF after cleaning and matching with other surveys as necessary. Two large clusters are observed in $A$-$\gamma$-space. Selection criteria are applied to this data for analysis of candidate QSOs, stars and galaxies. $\log(A)$ is in base 10 and based on $A$ units of in magnitudes.} 
    \label{fig:AllData} 
\end{figure}

\begin{figure}
	\centering
	\resizebox{\hsize}{!}{\includegraphics{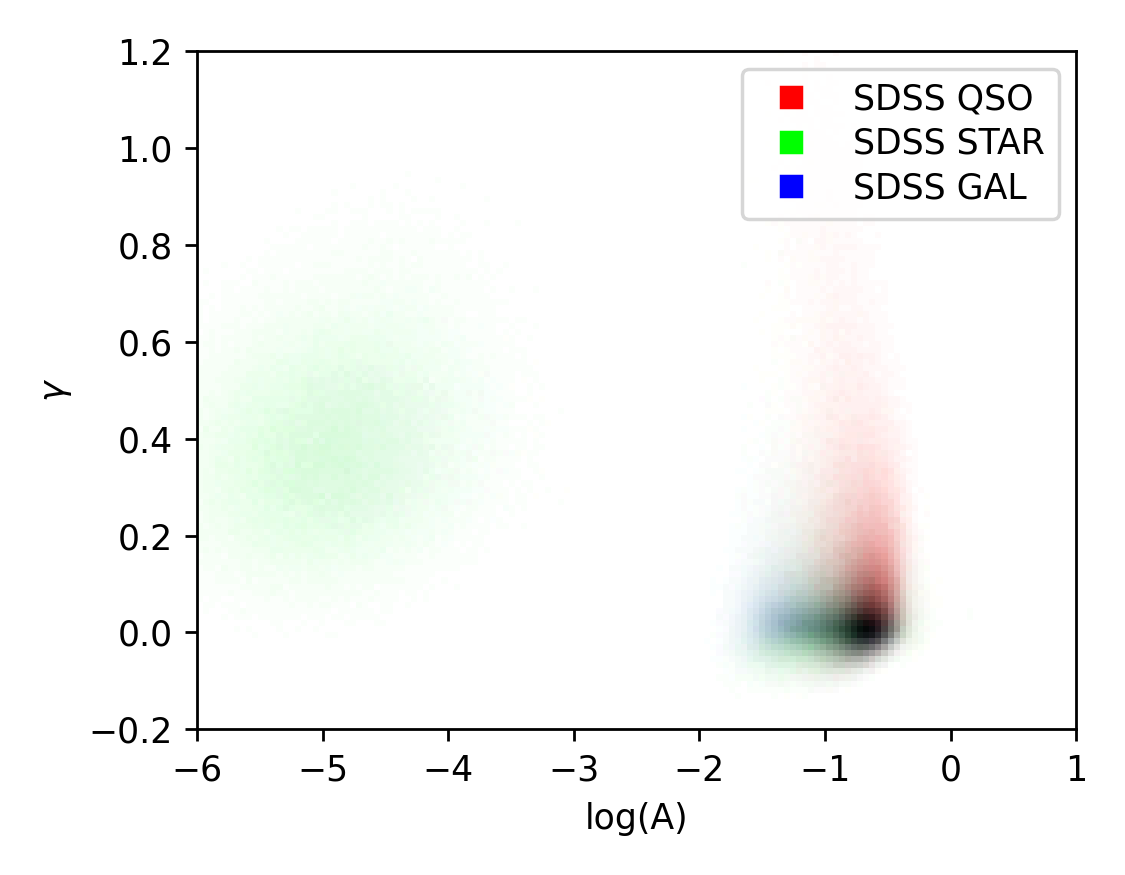}}
    \resizebox{\hsize}{!}{\includegraphics{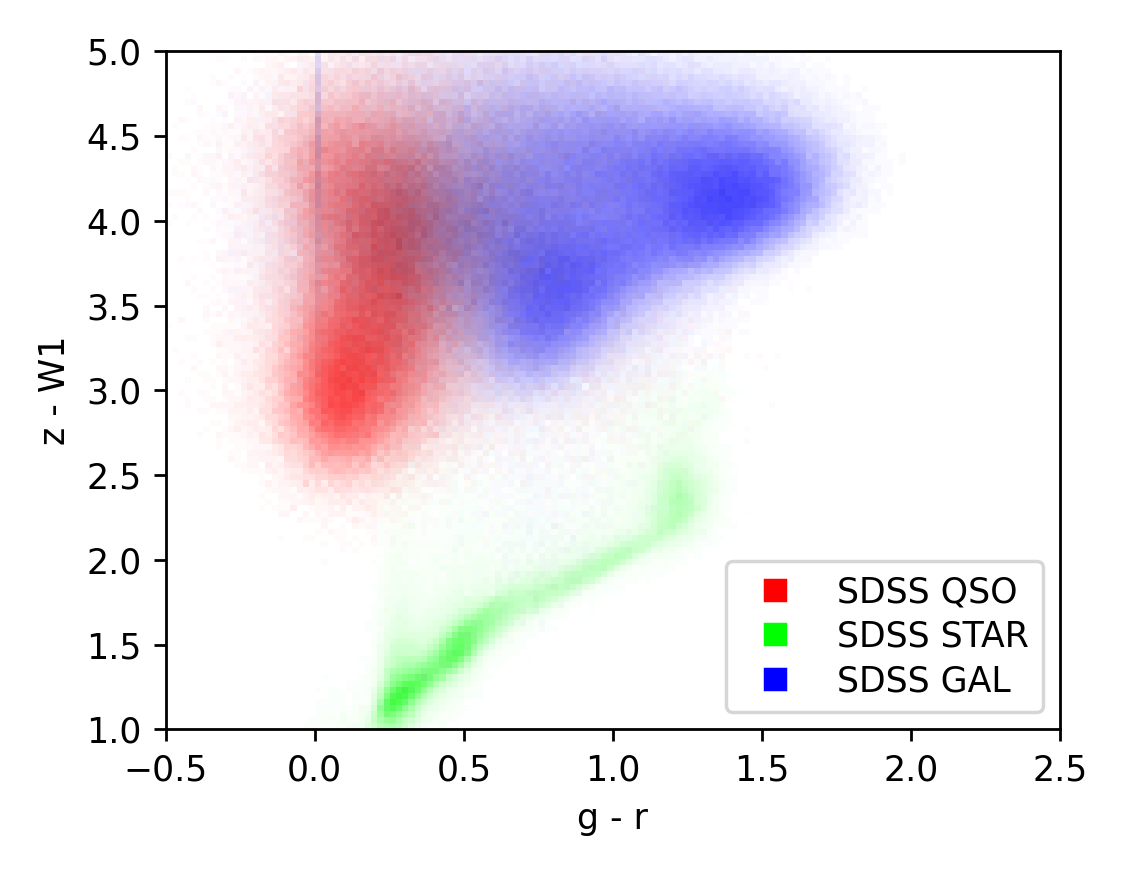}}
    \resizebox{\hsize}{!}{\includegraphics{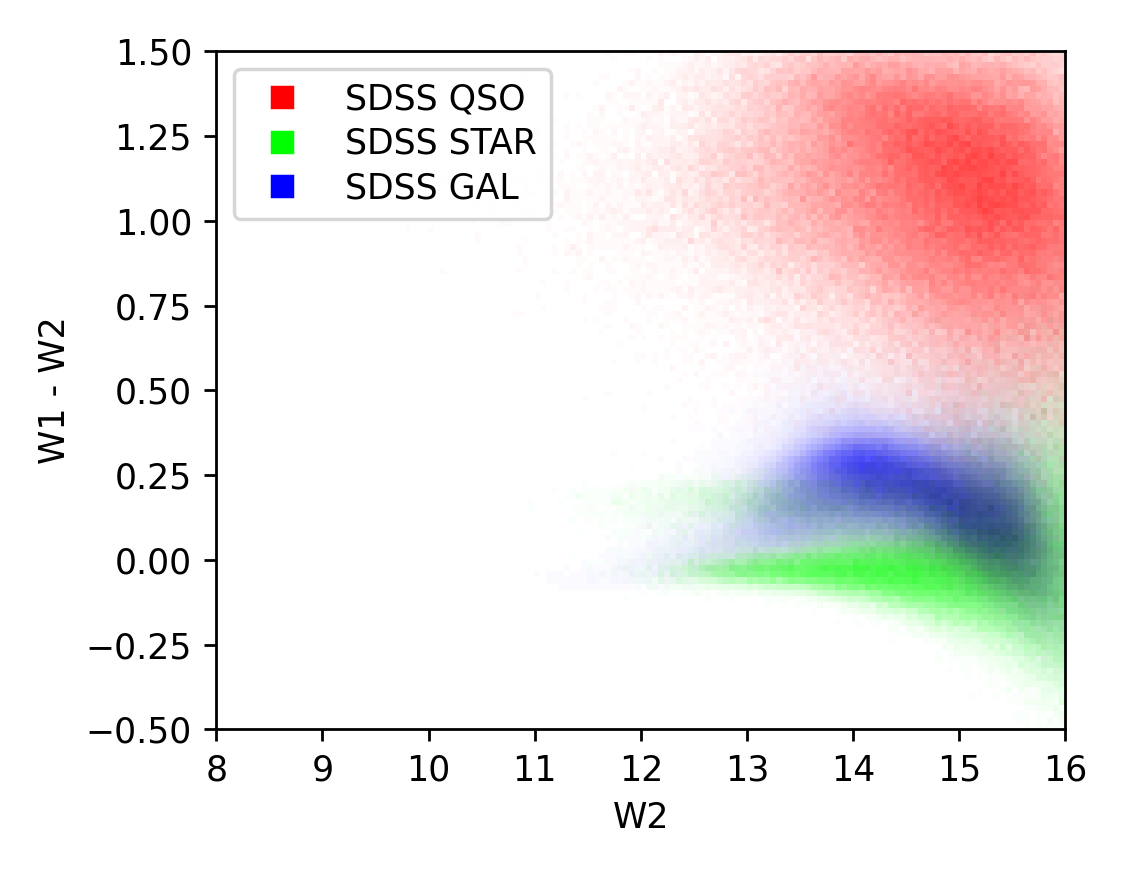}}
	\caption{All spectroscopically confirmed stars (green), QSOs (red) and galaxies (blue) from SDSS plotted in variability fit parameters (top) and colours: $g-r$ vs. $z-W1$ (middle) and $W2$ vs. $W1-W2$ (bottom). Marker colours show the object class and blend to grey or black when multiple classes occupy the same parameter region. A heatmap of all spectroscopically confirmed objects is found in Fig. \ref{fig:heatmaps_SDSS} 
	for comparison with Fig. \ref{fig:AllData}.}
    \label{fig:SDSSClasses} 
\end{figure}

Fig. \ref{fig:SDSSClasses} contains similar plots to Fig. \ref{fig:AllData}, but only for data with spectroscopic SDSS classes. 
In the variability parameter plot, this is 2.5~\% of the fitted sources (1\,748\,047), and the colour diagrams show the 2.276~\% (1\,613\,916) with matches in SDSS, WISE and PS1. 
The colours denote different classes: red for QSOs, green for stars and blue for galaxies. These are scaled based on relative density, with full saturation for areas of parameter space with one class and maximum relative density, white for areas without data and black or grey for areas with high relative densities of multiple classes. 
Note that this means a space that is more green than blue can still have more galaxies than stars. The full dataset includes more galaxies than stars and QSOs combined (see Sect. \ref{sec:data:SDSS}). 

In the top panel of $A$-$\gamma$ space, we notice the two clusters of Fig. \ref{fig:AllData}. The one at $\log(A)\sim-0.8$ has high relative densities of all classes, but they
are spread out along different axes in the $A$-$\gamma$ plane -- more galaxies have $\log(A)<-1$ and more QSOs have $\gamma>0.1$. Stars are mostly in the same areas as galaxies, but with more spread in $\gamma$ and higher relative density at $\log(A)<-3$. 
In the middle panel, the stellar locus is immediately recognisable, as well as the main distributions of QSOs and galaxies \citep{ansari,dustin}. The latter two include an overlap most apparent at $g-r\sim 0.4$ and $z-W1>0.3$.
The bottom panel shows that the three classes do have different distributions in $W1-W2$ vs. $W2$ too, but with a large overlap between stars and galaxies at $W2>15$.

\section{Photometric selection} \label{sec:photoclass}

We next explore how variability relates to broad band colours through photometric selections using colours, variability or both. 
Simple photometric selections of QSOs, stars and galaxies allow us to inspect differences between parameter distributions of photometrically and spectroscopically selected objects. 
Based on the distributions of Fig. \ref{fig:SDSSClasses}, we define regions of parameter space dominated by QSOs, stars or galaxies. 
\citet{schmidt} used a similar approach in $A$ and $\gamma$. For reliable selections, we focus on purity and accept a low completeness.
We avoid areas of high degeneracy between classes and achieve separate selection criteria for variability parameters and for colours.
All criteria are listed in Table \ref{tab:thresholds}. While these are simple, linear criteria in parameter space to study relations between colour and variability, more advanced techniques exist and have been deployed e.g. by \citet{ansari} in colour space. Such a study is presented in a companion paper \citep{paper2}.

We also examine properties of their QSO selection criteria, which were originally chosen by eye, for sources with large fit parameters compared to their uncertainties:
\begin{align}
    \frac{A}{\sigma_{A,-}}&> 2, &
    \frac{\gamma}{\sigma_{\gamma,-}} > 2. \label{eq:gammanot0}
\end{align}

\begin{table*}
      \centering
      \caption{Selection criteria.}
      \begin{tabular}{l l l l}
          \hline\hline
          Parameter & QSO candidates & Star candidates & Galaxy candidates \\
          \hline
          $\gamma$ & $>0.13$ & \ldots & $<0.1$ \\
          $\log_{10}(A)$ & $>0.11$ & $>0.01$ & $<0.1$ $\wedge$ $>0.01$ \\
          $g-r$ & $<0.2$ & \ldots & $>1$ \\
          $z-W1$ & $>2.8$ & $<2$ $\vee$ $<1 + 1.25 (g-r)$  & $> 2.3 + 1.25  (g-r)$ \\
          $W1-W2$ & $>0.75$ & $<-0.05$ & $>0.25$ $\wedge$ $<0.35$ \\
          $W2$ & \ldots & $<15$ & $<15$ \\
      \end{tabular}
      \label{tab:thresholds}
        \tablefoot{Criteria for selection of candidate QSOs, stars and galaxies. These are based on the SDSS distributions of Fig. \ref{fig:SDSSClasses} and the parameter regions are plotted in Fig. \ref{fig:heatmaps_SDSS}.}
\end{table*}{}

\subsection{Selection properties} \label{sec:performance}

The selection criteria of Table \ref{tab:thresholds} are based on purity and completeness of the labeled objects in Fig. \ref{fig:SDSSClasses} and then applied to the larger, unlabeled datasets of Fig. \ref{fig:AllData}. The criteria are illustrated in Fig. \ref{fig:heatmaps_SDSS} 
with those of \citet{schmidt} for comparison. 

Table \ref{tab:distribution} lists sample statistics of candidates selected within these parameter space regions of variability, colour or both. 
The Schmidt criteria are included for comparison. To estimate completeness, colour-based selections are given as percentages of the total set of objects with colour matches in WISE and PS1. 
Using both colour and variability criteria, the selections have low completeness and are very pure. However, selection bias in the spectroscopic sample affects purity estimates (see Sect. \ref{sec:specbias}). If we disregard the bias, a naive purity estimate for colour selected QSO candidates is 98.7~\%, assuming 15.5~\% of all PTF lightcurves with colour matches are QSOs. 
We get 99.3~\% purity for colour and variability selected QSOs, and 59.3
~\% with just variability criteria (assuming 15.3~\% QSOs in the latter case because WISE and PS1 matches are not required). 
Most candidates selected with colour and variability are registered with the same label as main type in SIMBAD; Fig. \ref{fig:SIMBAD_colvarclasses} shows statistics for each class for comparison. 

\textit{Gaia} DR3 \citep{Gaia,GaiaDR3} sources within one arcsecond of PTF sources have lower proper motions for QSO candidates. This is especially clear when colour information is included in QSO selection, as shown in Fig. \ref{fig:GaiaPM}. Objects with stellar colours and variability have the highest proper motions, as expected. 
Assuming a maximum of one \textit{Gaia} match per source, only 3.7\% of SDSS galaxies have a match in \textit{Gaia} with a measured proper motion, and the fraction decreases to 0.5\% for galaxy candidates selected by colour and variability. 
Further details are given in Appendix \ref{app:PM}.

If instead we use the Schmidt criteria and require the sources to have $A$ and $\gamma$ significantly different from zero (Eq. \ref{eq:gammanot0}), we find the most common SDSS class to be QSOs. However, these criteria lead to more contamination from stars and galaxies than the QSO criteria of Table \ref{tab:thresholds}. 

\subsubsection{Colour selection}
Fig. \ref{fig:ColorClasses} shows distributions of colour-based selections from the large dataset of fitted sources in Fig. \ref{fig:AllData}. The overall pattern is the same as in Fig. \ref{fig:SDSSClasses}, but the stars are more spread out. More of them have very low $A$ or high $\gamma$, and relatively fewer are found at $\log(A)\sim-0.8$ and $\gamma\sim0$.
According to Table \ref{tab:distribution}, the colour selected candidates have similar, high purities of $98.6-99.7$~\% for all classes, but with varying completeness -- 6.82~\% for galaxies, 7.92~\% for stars and 38.3~\% for QSOs.

\subsubsection{Variability selection}
\begin{figure}
	\centering
    \resizebox{\hsize}{!}{\includegraphics{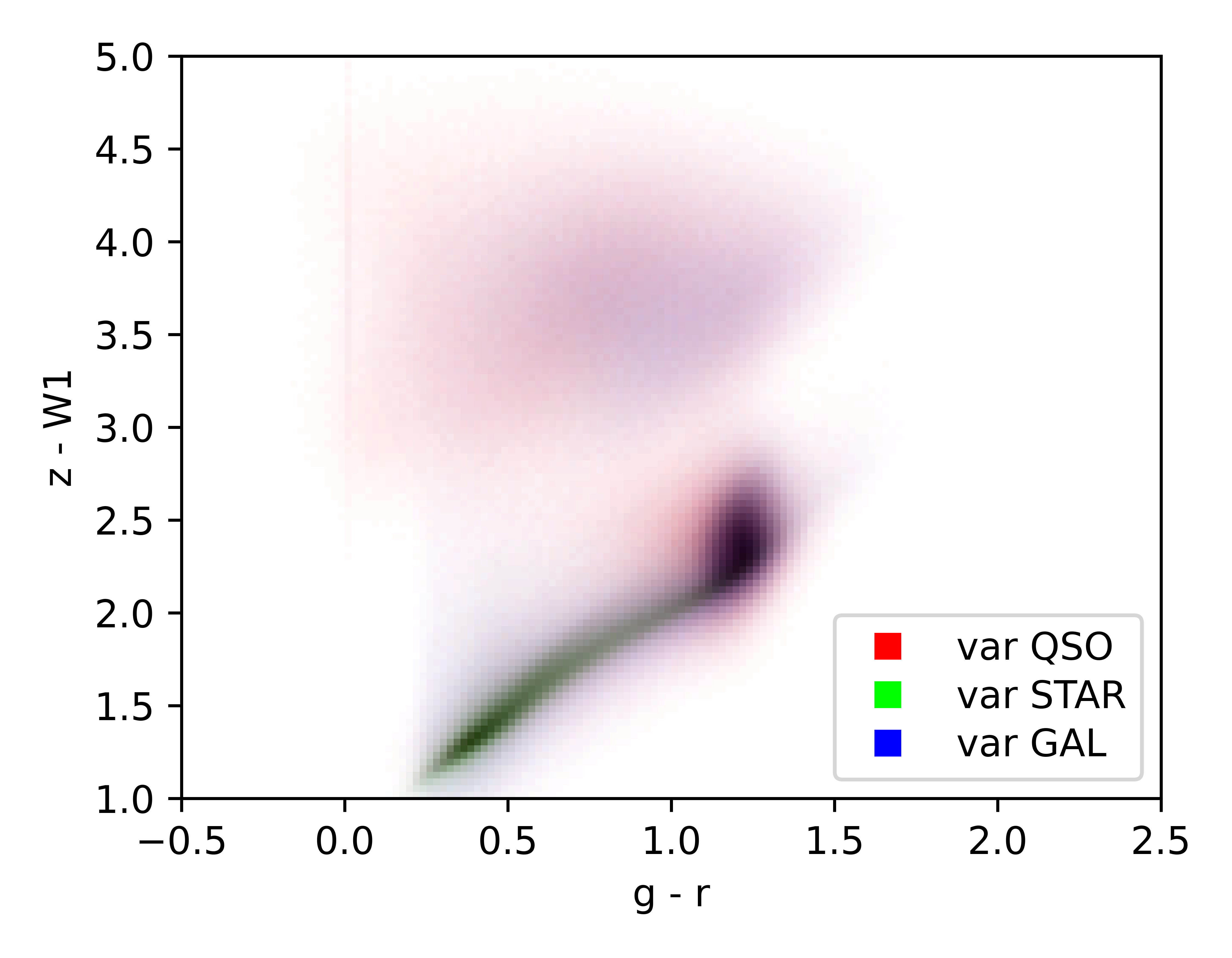}}
    \resizebox{\hsize}{!}{\includegraphics{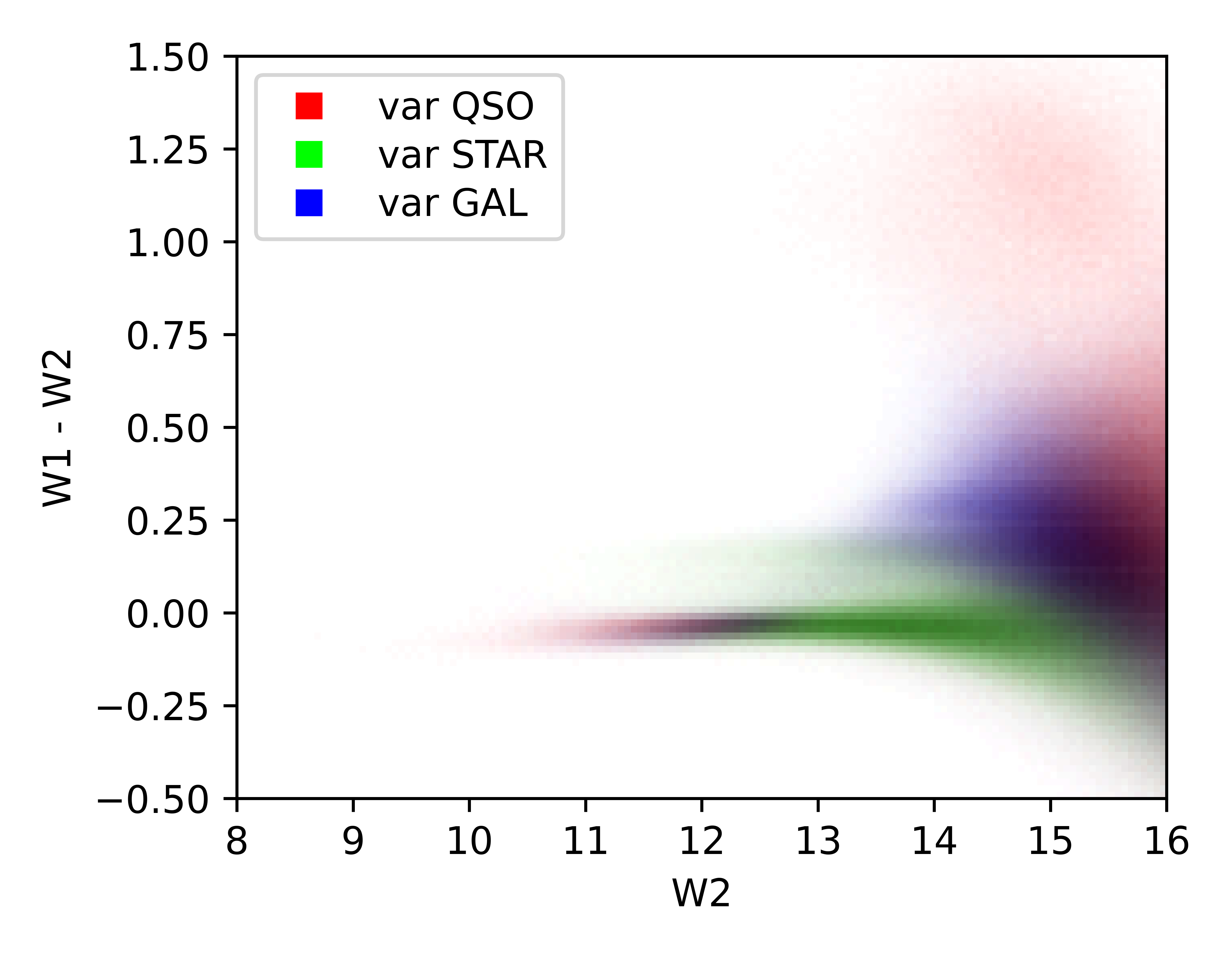}}
	\caption{Colour distributions of variability-selected objects from Fig. \ref{fig:AllData}. These fulfil the variability criteria of Table \ref{tab:thresholds} based on the upper diagram of in Fig. \ref{fig:SDSSClasses}.} 
    \label{fig:VarClasses} 
\end{figure}

We plot the variability-selected sources in Fig. \ref{fig:VarClasses}. 
Relying solely on $A$ and $\gamma$ is challenging according to the SDSS labels. For example, while the variability criteria for stars catch 53.4~\% of stars and 19.8
~\% of galaxies, they select more galaxies due to different population sizes. Naturally, criteria based on overlapping classes in $A$-$\gamma$-space lead to overlapping selections in colour space.

\subsection{Ambiguous sources} \label{sec:ambi}

Given the purity of the photometric selections in Table \ref{tab:distribution}, the contaminating sources are expected to have a high rate of misclassifications by SDSS. Checking the most confident photometric predictions with differing spectroscopic classifications, we find examples of spectra appearing more typical of 
the photometric candidate class. 
We inspect the spectra of the 32 photometric QSO candidates with stellar spectroscopic classifications, and judge $\sim 20$~\% to be QSOs and $\sim 50$~\% as possible QSOs. 
16 of the objects are in SIMBAD, registered as eight QSOs, four BL Lacertae objects, one blazar and just three stars. 
For example, SDSS J120429.34+495814.4, with the spectrum of Fig. \ref{fig:QSO_spectrum}, has the characteristic broad lines of a QSO. This object is part of the Data Release 12 Quasar catalog from SDSS \citep{DR12Q}. However, it is spectroscopically confirmed as a star in SDSS DR17. 
This misclassification analysis shows the power of the photometric selection criteria of this work, although we do not expect high rates of misclassifications in the full SDSS spectroscopically confirmed sample.

Variable galaxies in the Table \ref{tab:thresholds} QSO region of $A$ and $\gamma$ have higher redshifts, $W2$ and $z-W1$ than other galaxies (typically redshift 0.5 vs. 0.1, see Figs. \ref{fig:GAL_redshifts} and \ref{fig:GAL_var})
. In SIMBAD, variable galaxies are more often registered as the brightest galaxy in a cluster than other galaxies are. The fraction increases from 9 to 15~\%. There is also a low, but relatively much higher, fraction of radio sources which goes from 1.6 to 3.7~\%.  SDSS QSOs have similar redshifts independently of variability, except more non-QSO-like variability at $z<0.3$ (see Fig. \ref{fig:GAL_redshifts}).

Spectroscopic QSOs variability-selected as stars or galaxies are more often found at $W1-W2<0.75$ and $W2<14$ than other QSOs. This difference in WISE colours is smaller than for spectroscopic galaxies and stars in different photometric classes, though, as illustrated in Figs. \ref{fig:GAL_var} -- \ref{fig:STAR_var}.

\begin{figure}
	\centering
    \resizebox{\hsize}{!}{\includegraphics{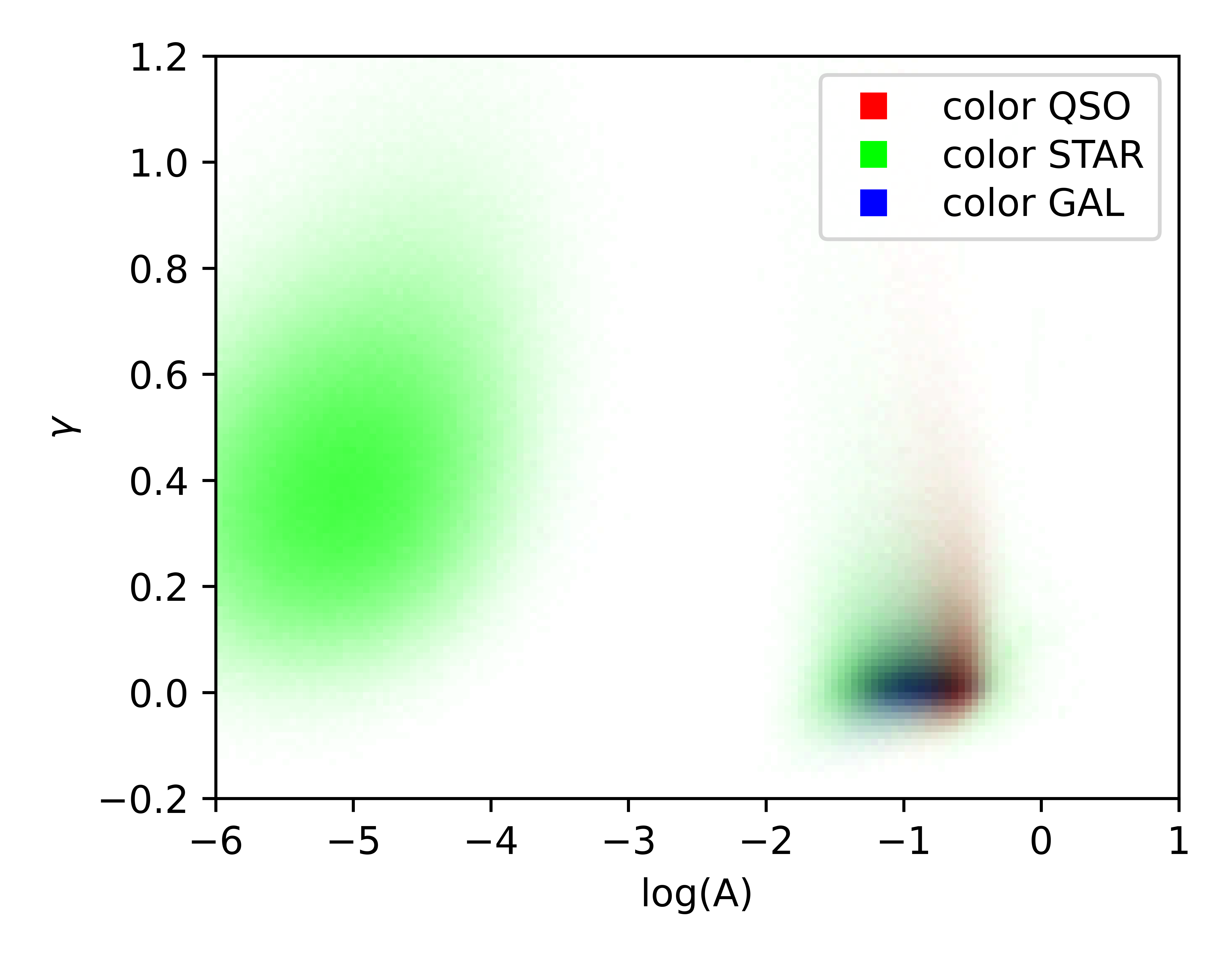}}

	\caption{Variability of objects of Fig. \ref{fig:AllData} selected by the colour criteria from Table \ref{tab:thresholds}. These are based on the colour distributions of Fig. \ref{fig:SDSSClasses}. }
    \label{fig:ColorClasses}
\end{figure}

\begin{table*}
      \centering
      \caption{Selection statistics.}
      \begin{tabular}{l l l l l}
          \hline\hline
          Selection & QSOs & Stars & Galaxies & All\\
          \hline
          All & 268\,230 & 389\,317 & 1\,090\,500  & 70\,920\,904 \\  
         \hspace{0.5cm} Relative frequency in full sample  & 0.3782~\% & 0.5489~\% & 1.538~\% & \ldots \\
         \hspace{0.5cm} Relative frequency in spec sample  & 15.34~\% & 22.27~\% & 62.38~\% & \ldots\\
          \hline
          Colour and variability QSO candidate & 31\,404 & 32 & 191 & 70\,503 \\
          \hspace{0.5cm}  Completeness & 12.53~\% & 0.011~\% & 0.018~\% & 0.1237~\% \\
          \hspace{0.5cm}    Purity & 99.3~\% & 0.10~\% & 0.60~\% & \ldots \\
          \hline
          Colour and variability star candidate &  15 & 21\,129 & 194 & 2\,995\,546 \\
          \hspace{0.5cm}  Completeness &  0.006~\% & 7.24~\% & 0.018~\% & 5.256~\% \\
          \hspace{0.5cm}    Purity & 0.07~\% & 99~\% & 0.91~\% & \ldots \\
          \hline
          Colour and variability galaxy candidate & 28 & 25  & 28\,076 & 107\,227 \\
          \hspace{0.5cm}  Completeness & 0.011~\% & 0.009~\% & 2.62~\% & 0.1881~\% \\
          \hspace{0.5cm}    Purity & 0.10~\% & 0.09~\% & 99.8~\% & \ldots \\
          \hline
          Variability QSO candidate & 75\,690 & 12\,704 & 39\,284 & 3\,390\,617 \\ 
          \hspace{0.5cm}  Completeness & 28.2~\% &  3.26~\% & 3.60~\% & 4.781~\% \\ 
          \hspace{0.5cm}    Purity & 59.3~\% & 9.95~\% & 30.8~\% & \ldots \\
          \hline
          Variability star candidate & 14\,005 & 207\,701  & 216\,312 & 22\,854\,151 \\
          \hspace{0.5cm}  Completeness & 5.22~\% & 53.4~\% & 19.84~\% & 32.225~\% \\
          \hspace{0.5cm}    Purity & 3.20~\% & 47.4~\% & 49.4~\% & \ldots \\
          \hline
          Variability galaxy candidate & 12\,470  & 44\,170  & 228\,634 &  9\,140\,797 \\
          \hspace{0.5cm}  Completeness & 4.65~\% & 11.35~\% & 20.97~\% & 12.889~\% \\
          \hspace{0.5cm}    Purity & 4.37~\% & 15.48~\% & 80.1~\% & \ldots \\
          
          \hline
          Colour QSO candidate & 96\,050 & 237 & 1\,045 & 277\,394 \\ 
          \hspace{0.5cm} Completeness & 38.3~\% & 0.081~\% & 0.098~\% & 0.4867~\% \\
          \hspace{0.5cm}    Purity & 98.7~\% & 0.24~\%  & 1.07~\% & \ldots \\
          \hline
          Colour star candidate & 17 & 23\,111 & 319 & 3\,495\,123 \\
          \hspace{0.5cm}  Completeness &  0.007~\% & 7.92~\% & 0.030~\% & 6.133~\% \\
          \hspace{0.5cm}    Purity & 0.07~\% & 98.6~\% & 1.36~\% & \ldots \\
          \hline
          Colour galaxy candidate & 152 & 73 & 73\,112 & 322\,126 \\
          \hspace{0.5cm}  Completeness & 0.061~\% & 0.025~\% & 6.82~\% & 5.65~\% \\
          \hspace{0.5cm}    Purity & 0.21~\% & 0.10~\% & 99.7~\% & \ldots \\
          
          \hline
          Schmidt region & 85\,012 & 24\,576 & 79\,279 & 5\,484\,750  \\
            \hspace{0.5cm}  Completeness & 31.7~\% & 6.31~\% & 7.27~\% & 7.734~\% \\
            \hspace{0.5cm}  Purity & 45.0~\% & 13.01~\% & 42.0~\% & \ldots \\
          \hline
      \end{tabular}
      \label{tab:distribution}
        \tablefoot{
        Sample statistics of selections in variability and colour, including counts, completeness and purity (see Eq. \ref{eq:performance}). 
        In row 1 (All), we compare spectroscopically classified objects to the full fitted source count and to the full set with spectroscopic classes to get relative frequencies in each set.
        In row 5-7 and 11, for computing purity and completeness, selection counts are compared to the total spectroscopic counts. In row 2-4 and 8-10, comparisons also require matches in WISE and PS1, since colours are used.
        }
  \end{table*}{}
  
\section{Discussion} \label{sec:discussion}
QSOs, stars and galaxies are distributed differently in $A$ and $\gamma$ and in colours, but with overlaps limiting the quality of simple selections. We see this in differences between the SDSS class distributions of Fig. \ref{fig:SDSSClasses} and the corresponding candidate class plots in Fig. \ref{fig:ColorClasses} and \ref{fig:VarClasses} based on colour and variability, respectively. The overall patterns can still be recognised. 
Class information in the SF variability parameters is especially interesting for objects without spectroscopy and limited colour measurements. Variability alone does not give as reliable candidates as colour selection, but colour and variability combined can break degeneracies and select classes better than separately. 

\subsection{$A$-$\gamma$ clusters}

In $A$-$\gamma$ space, we see two large clusters. One of them is likely an artefact from objects with undetectable variability, leading the MCMC to suggest arbitrarily small $A$ and a large range of $\gamma$ values. There is only one clear cluster at higher $A$ ($\log(A)\sim-0.8$), but the QSOs and galaxies are spread along different axes. This is most apparent in Fig. \ref{fig:SDSSClasses}. 
The galaxies cover a broad range of $A$, but their variability is rarely timescale dependent, at least not on most relevant scales. This is shown by the low $\gamma$ values. In contrast, QSOs are even more variable (high $A$) with a clearer timescale dependence. 
Stars are spread out more evenly in $A$ and $\gamma$, limiting selection purity and reflecting the diverse nature of stellar variability.

\subsection{Comparison with SDSS lightcurve analysis} 

The selection criteria of Table \ref{tab:thresholds} are simple and focused on purity. For QSOs, they differ from those by \citet{schmidt}. 
Table \ref{tab:distribution} shows a purer QSO set with slightly lower completeness, than if we apply the Schmidt criteria. 
This may be due to slightly different fitting and outlier removal or differences in noise and measurements between PTF and SDSS. 
The dataset presented in this paper is more representative of all observed object types, as it includes all sources from the PTF survey that pass through the data cleaning of Sect. \ref{sec:cleaning}, whereas \citet{schmidt} introduced specific types of contaminants and in specific ratios. 

\subsection{Spectroscopic selection bias} \label{sec:specbias}

When only 2.5~\% of sources have spectroscopic classifications, it shows a need for other methods for identification. It also allows for significant selection bias in sources with SDSS spectra compared to sources with long PTF lightcurves. This affects purity estimates, and the choice and evaluation of selection regions. 

There are differences in the distributions of sources with and without spectroscopic classes. This is especially clear if we compare Fig. \ref{fig:AllData} to Fig. \ref{fig:heatmaps_SDSS}.
For example, SDSS is missing spectra for objects at low $W2$, including a band of sources at low $W1-W2$. At $10<W2<12$ and $-0.5<W2-W1<-0.25$ we have a very mixed group, with sources typically marked as stars (21~\%) or binary star systems in SIMBAD. 
 
SDSS also has a bias in favour of sources with $z-W1>3$. This is part of the reason why in Fig. \ref{fig:VarClasses}, QSOs and galaxies are relatively infrequent at those values compared to Fig. \ref{fig:SDSSClasses}. If more objects are included in star-dominated areas, they also include a larger fraction of the galaxies and QSOs. Many of these are actually stars, though, but wrongly selected by variability. Especially variable stars late in the main sequence can be difficult to distinguish from QSOs and galaxies; the parameter region at $2.1<z-W1<2.6$ and $1.1<g-r<1.4$ is dominated by stars for all three variability selections. Variability-selected star candidates in the area are at least 86~\% stars judging by the most common stellar classifications registered in SIMBAD ("Star", "low-mass*" and "PM*"). Galaxy and QSO candidates are at least 79~\% and 72~\% stars, respectively, showing a small difference in the nature of these objects. For spectroscopic stars, those with QSO- or galaxy-like variability are more spread out in $z-W1$ and found at higher values of $W1-W2$. This is illustrated in Fig. \ref{fig:STAR_var}. The variability does indicate a physical difference in these cases. 

In $W2$ vs. $W1-W2$, the spectroscopic classes overlap at $W2>15$, and the variability selections overlap even more. Objects  
at $-0.25<W2-W1<0$ and $10<W2<12$ are mostly registered as stars in SIMBAD, and SDSS does not classify any of them as QSOs, although many have QSO-like variability.

SDSS matched sources have relatively longer time spans and more epochs per lightcurve, as shown in Fig. \ref{fig:epochs_per_lc}. 
Hence, variability estimation of the full sample might be less accurate -- spreading out sources in $A$-$\gamma$ space. Therefore, removing the most sparsely sampled sources is important. To both include large data sets and be confident in the results, the balance of data cleaning  
will also be important in future surveys such as the 
LSST. Even for sources with >100 epochs over >5 years , the spectroscopic classes still have an overlap at $\log A \sim -0.8$ and $\gamma \sim 0$, but it is about 50~\% smaller in both $\log A$ and $\gamma$. Longer timescales may change the selected populations, for example by increasing the fraction of type II to type I AGN \citep{cicco2015,cicco2019}.

\subsection{Photometric selection bias} \label{sec:photobias}

We select pure sets of each class, but with low completeness and a bias for sources with specific parameters.
With variability, we only select stars with low $A$ -- but we know variable stars exist. They are just difficult to isolate, and so 
reduce completeness for stars and purity for galaxies and QSOs. 
Type I and Type II AGN differ in colours and SFs, and so are not best selected by one simple set of criteria either \citep{cicco2022}.
The most densely populated variability region, at $\log A \sim -0.8$ and $\gamma \sim 0$, is not covered by the criteria at all. The same goes for the dense colour diagram area at high $W2$. 
In Fig. \ref{fig:SDSSClasses}, galaxies dominate a triangular area of $g-r$ vs. $z-W1$ with two large clusters. The criteria only cover one of them. In SIMBAD, the cluster at high $g-r$ has more galaxies labeled as being part of a cluster, and especially as the brightest galaxy in a cluster. A more advanced selection method could solve these issues \citep{paper2}. 

The surveys are not completely representative of the sky, which is not observed uniformly (see Fig. \ref{fig:Nobj_per_patch}). 
We include fewer stars at low Galactic latitudes, where Galactic extinction affects colours more and there is a higher risk of mismatches with nearby sources.
Stars have the fewest colour matches, indicating an under-representation in WISE or PS1. 
A change in ratios of object types and stellar subtypes affects purity. The reason is that the number of true or false positives depends on the selection of objects that are not equally difficult to distinguish from or as stars. E.g. including more variable stars could lower the QSO selection purity -- both due to the overlap with QSO variability and the increased prior probability of a classified object being a star. However, including more data would also mean there is more data to learn from. The balance of object type frequencies is relevant if we apply Table \ref{tab:thresholds} criteria to other datasets, and in evaluation of criteria of \citet{schmidt} on PTF data. Completeness is computed independently for each class, but could also be affected by a change in the fraction of subtypes.

\subsection{Perspectives}

If we were to base selection criteria on distributions from SDSS confirmed objects and evaluate on the same set, we would be overfitting. However, here the goal is only to estimate class distributions and relations between variability and colour. Machine learning could automatically classify all objects based on the SDSS labeled subset. This would allow us to select more sources and examine probabilities of belonging to each class. It would quantify how variability breaks degeneracies and improves selections by colours and magnitudes. This will be performed in paper two of VILLAIN \citep{paper2}, including a table of all variability parameters and classifications.
Accurate photometric selections can identify new QSO candidates and prepare us for analysis of large surveys like the LSST. Optical variability of galaxies, including in the PTF R-band, is also known be useful for identifying AGN missed by other techniques \citep{lowmassAGN}. It would be interesting to study subtypes, such as by variability and redshift differences of type I and II AGN, like \citet{cicco2022}. Intermediate-redshift QSOs can have colours comparable to those of stars, so variability could be more valuable for those \citep{stellarcolQSO}. 
Another prospect is selecting non-variable stars for photometric calibration or for a homogeneous study of variability across stellar subtypes.

We have assumed that the objects have simple power law variability, but this is not necessarily a good model for all sources or on all timescales.   
One could fit e.g. exponential or DRW models instead and analyse the differences, though we expect the overall selections and challenges to be similar. Advanced models can capture more variability information but require more resources \citep{SFgaps}. More parameters could be used for selection, such as proper motions or photometric redshifts.

\section{Conclusion} \label{sec:conclusion}

We have devised a procedure for the homogeneous analysis of 71 million PTF lightcurves. We have fitted them with joint-likelihood SF models and studied regions in both variability and colour space. SF power law variability is most useful outside the $\log(A)\sim-0.8$ and $\gamma\sim0$ region. 
We select photometric sets of 99.3~\% spectroscopic purity for QSOs, 99~\% for stars and 99.8~\% for galaxies. The spectroscopic classifications are, however, wrong for 20 -- 50~\% of objects photometrically identified as QSOs but spectroscopically as stars. 
The large PTF sample allows us to discover these rare cases and assemble a set of 31\,404 QSO candidates by colour and variability. With only variability, spectroscopic purity drops to 59.3~\% and with only colour, it is 98.7~\%.

Using SF joint likelihoods on the entire PTF survey, we have shown how variability might be used on future large datasets including the  
LSST. When new measurements are added to a lightcurve, complete reprocessing can be avoided, as likelihood information on the previous segment of the lightcurve is already stored in $A$ and $\gamma$. In a survey with the LSST foreseen depth, a colour-plus-variability method can provide a large sample of faint astrometric standards for the internal calibration of extremely large telescopes, which require objects beyond the depth of \textit{Gaia}.

In each survey, and depending on computational resources, one must balance sample size and fitting accuracy via data cleaning and selection methods. The value of variability and colours depend on the survey and sources, but in general for PTF, cross-matching colours should be prioritised. 

Considering simple cuts in both variability and colour, the completeness is at 12.5
~\%, so a machine learning method that balances purity and completeness has potential for creating larger QSO samples for studying e.g. cosmology. This is examined in the companion VILLAIN paper \citep{paper2}, where we also release a table of classifications and parameters for all fitted PTF sources. Such a large data-set would allow e.g. a complementary selection of rare objects, such as lensed QSOs.

\begin{acknowledgements} 

    This work was supported by a Villum Investigator grant (project number 16599).
    SHB was also supported by a grant from the Danish National Research Foundation.
    AA was also supported by a Villum Experiment grant (project number 36225). 

    This publication makes use of data products from the Wide-field Infrared Survey Explorer, which is a joint project of the University of California, Los Angeles, and the Jet Propulsion Laboratory/California Institute of Technology, funded by the National Aeronautics and Space Administration.

    The Pan-STARRS1 Surveys were made possible through contributions by
    the Institute for Astronomy, the University of Hawaii, the Pan-STARRS
    Project Office, the Max-Planck Society and its participating
    institutes, the Max Planck Institute for Astronomy, Heidelberg and the
    Max Planck Institute for Extraterrestrial Physics, Garching, The Johns
    Hopkins University, Durham University, the University of Edinburgh,
    the Queen's University Belfast, the Harvard-Smithsonian Center for
    Astrophysics, the Las Cumbres Observatory Global Telescope Network
    Incorporated, the National Central University of Taiwan, the Space
    Telescope Science Institute, and the National Aeronautics and Space
    Administration under Grant No. NNX08AR22G issued through the Planetary
    Science Division of the NASA Science Mission Directorate, the National
    Science Foundation Grant No. AST-1238877, the University of Maryland,
    Eotvos Lorand University (ELTE), and the Los Alamos National
    Laboratory.The Pan-STARRS1 Surveys are archived at the Space Telescope
    Science Institute (STScI) and can be accessed through MAST, the
    Mikulski Archive for Space Telescopes. Additional support for the
    Pan-STARRS1 public science archive is provided by the Gordon and Betty
    Moore Foundation.

    The Pan-STARRS1 Surveys, Chambers, K.C., et al. 2016, arXiv:1612.05560;
    Pan-STARRS Data Processing System,
      Magnier, E. A., et al. 2016, arXiv:1612.05240;
    Pan-STARRS Pixel Processing: Detrending, Warping, Stacking,
      Waters, C. Z., et al. 2016, arXiv:1612.05245;
    Pan-STARRS Pixel Analysis: Source Detection and Characterization,
      Magnier, E. A., et al. 2016, arXiv:1612.05244;
    Pan-STARRS Photometric and Astrometric Calibration,
      Magnier, E. A., et al. 2016, arXiv:1612.05242;
    The Pan-STARRS1 Database and Data Products,
      Flewelling, H. A., et al. 2016, arXiv:1612.05243.
  
    Funding for the Sloan Digital Sky 
    Survey IV has been provided by the 
    Alfred P. Sloan Foundation, the U.S. 
    Department of Energy Office of 
    Science, and the Participating 
    Institutions. 

    SDSS-IV acknowledges support and 
    resources from the Center for High 
    Performance Computing  at the 
    University of Utah. The SDSS 
    website is www.sdss.org.
    
    SDSS-IV is managed by the 
    Astrophysical Research Consortium 
    for the Participating Institutions 
    of the SDSS Collaboration including 
    the Brazilian Participation Group, 
    the Carnegie Institution for Science, 
    Carnegie Mellon University, Center for 
    Astrophysics | Harvard \& 
    Smithsonian, the Chilean Participation 
    Group, the French Participation Group, 
    Instituto de Astrof\'isica de 
    Canarias, The Johns Hopkins 
    University, Kavli Institute for the 
    Physics and Mathematics of the 
    Universe (IPMU) / University of 
    Tokyo, the Korean Participation Group, 
    Lawrence Berkeley National Laboratory, 
    Leibniz Institut f\"ur Astrophysik 
    Potsdam (AIP),  Max-Planck-Institut 
    f\"ur Astronomie (MPIA Heidelberg), 
    Max-Planck-Institut f\"ur 
    Astrophysik (MPA Garching), 
    Max-Planck-Institut f\"ur 
    Extraterrestrische Physik (MPE), 
    National Astronomical Observatories of 
    China, New Mexico State University, 
    New York University, University of 
    Notre Dame, Observat\'ario 
    Nacional / MCTI, The Ohio State 
    University, Pennsylvania State 
    University, Shanghai 
    Astronomical Observatory, United 
    Kingdom Participation Group, 
    Universidad Nacional Aut\'onoma 
    de M\'exico, University of Arizona, 
    University of Colorado Boulder, 
    University of Oxford, University of 
    Portsmouth, University of Utah, 
    University of Virginia, University 
    of Washington, University of 
    Wisconsin, Vanderbilt University, 
    and Yale University.
    
    This research made use of the cross-match service provided by CDS, Strasbourg. 
    
    This research made use of Astropy,\footnote{http://www.astropy.org} a community-developed core Python package for Astronomy (\citeauthor{Astropy1}, \citeyear{Astropy1}; \citeauthor{Astropy2}, \citeyear{Astropy2}). 
    
    This research has made use of the NASA/IPAC Infrared Science Archive, which is funded by the National Aeronautics and Space Administration and operated by the California Institute of Technology.
    
    This research makes use of the SciServer science platform (www.sciserver.org).

    SciServer is a collaborative research environment for large-scale data-driven science. It is being developed at, and administered by, the Institute for Data Intensive Engineering and Science at Johns Hopkins University. SciServer is funded by the National Science Foundation through the Data Infrastructure Building Blocks (DIBBs) program and others, as well as by the Alfred P. Sloan Foundation and the Gordon and Betty Moore Foundation.
    
    This research has made use of the VizieR catalogue access tool, CDS, Strasbourg, France (DOI : 10.26093/cds/vizier). The original description of the VizieR service was published in 2000, A\&AS 143, 23.
    
    This work has made use of data from the European Space Agency (ESA) mission
    {\it Gaia} (\url{https://www.cosmos.esa.int/gaia}), processed by the {\it Gaia}
    Data Processing and Analysis Consortium (DPAC,
    \url{https://www.cosmos.esa.int/web/gaia/dpac/consortium}). Funding for the DPAC
    has been provided by national institutions, in particular the institutions
    participating in the {\it Gaia} Multilateral Agreement.
    
\end{acknowledgements}


\bibliographystyle{aa} 
\bibliography{bibliography.bib}

\clearpage
\begin{appendix}
\section{Sky maps} \label{app:sky}
\begin{figure}[H]
	\centering
	\resizebox{\hsize}{!}{\includegraphics{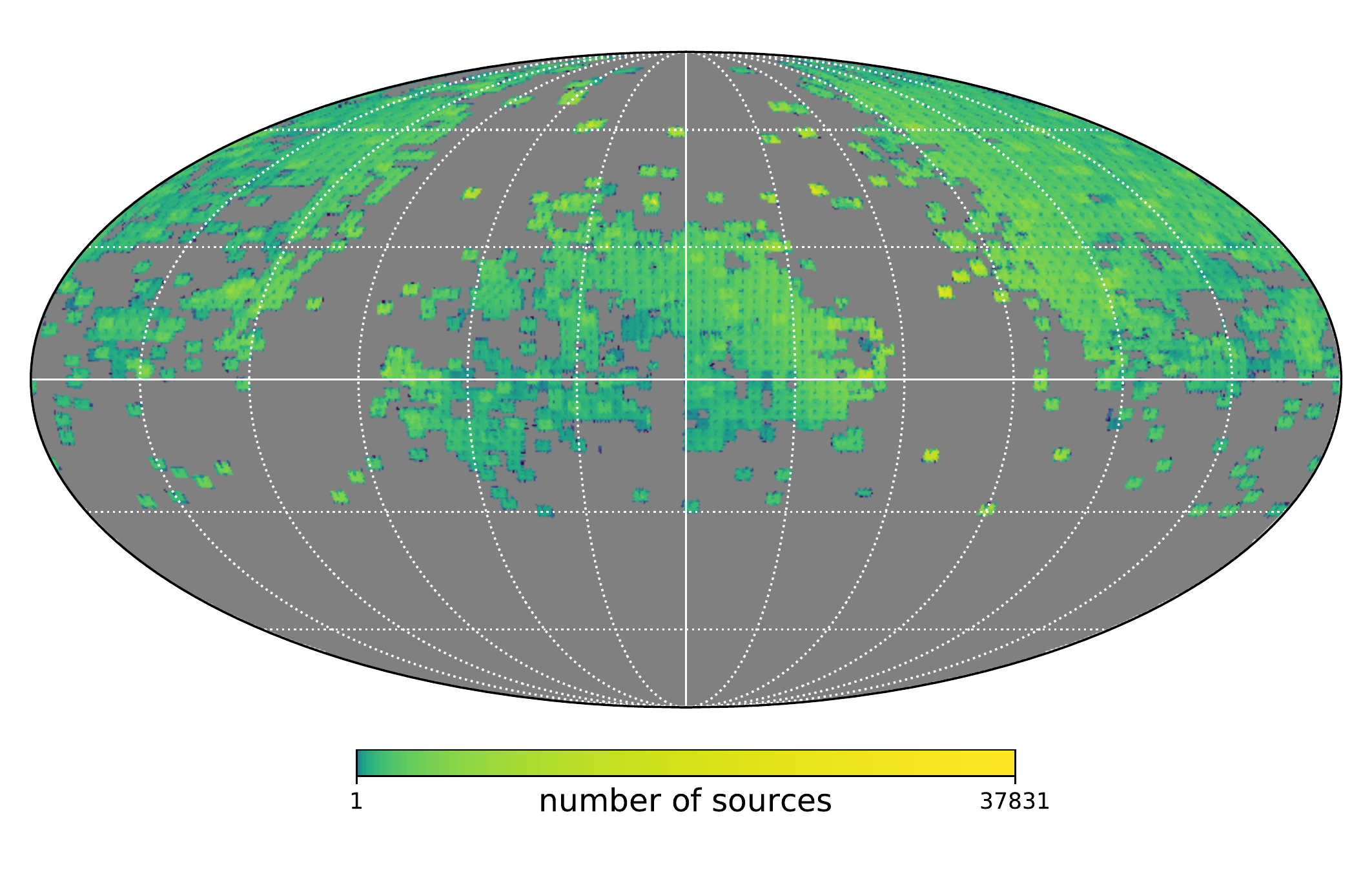}}
	\resizebox{\hsize}{!}{\includegraphics{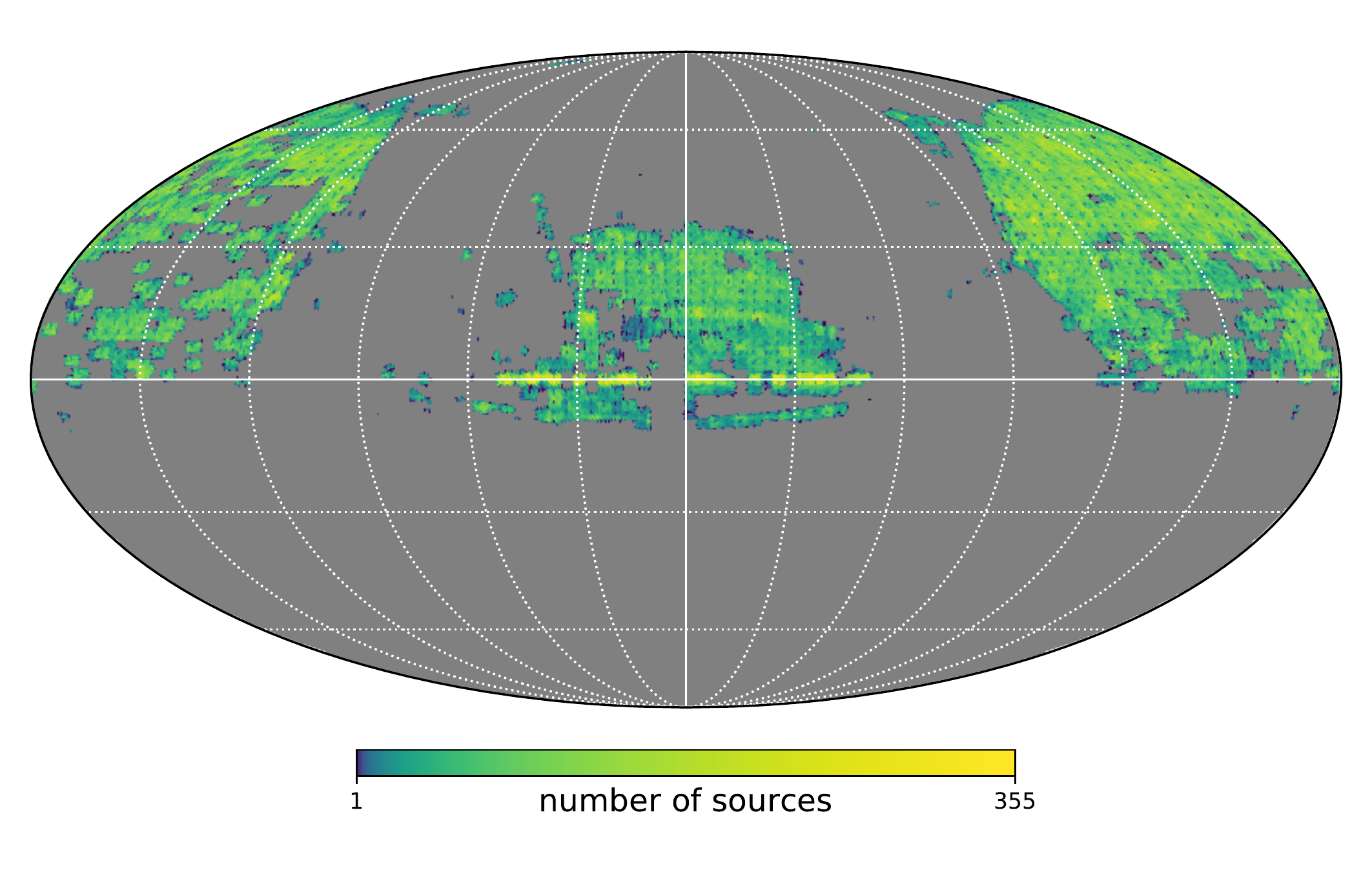}}
	\caption{Sky distributions of fitted and accepted PTF lightcurves (top) and of those also matched to  spectroscopic classifications in SDSS (bottom). The coordinates are equatorial, with RA increasing to the right. 
	In grey areas, no data exists or remains after the data cleaning of Sect. \ref{sec:cleaning}.} 
    \label{fig:Nobj_per_patch}
\end{figure}

The sources with PTF lightcurves are mainly from the northern hemisphere. These are distributed according to the top diagram of Fig. \ref{fig:Nobj_per_patch} after the data cleaning of Sect. \ref{sec:cleaning}. When we also match to SDSS spectroscopic classifications, the sources in the bottom diagram are left. The differences indicate a bias in photometry between the surveys, as mentioned in Sect. \ref{sec:photobias}.

\section{SDSS matched data} \label{app:SDSS}

\begin{figure}[H]
	\centering
    \resizebox{\hsize}{!}{\includegraphics{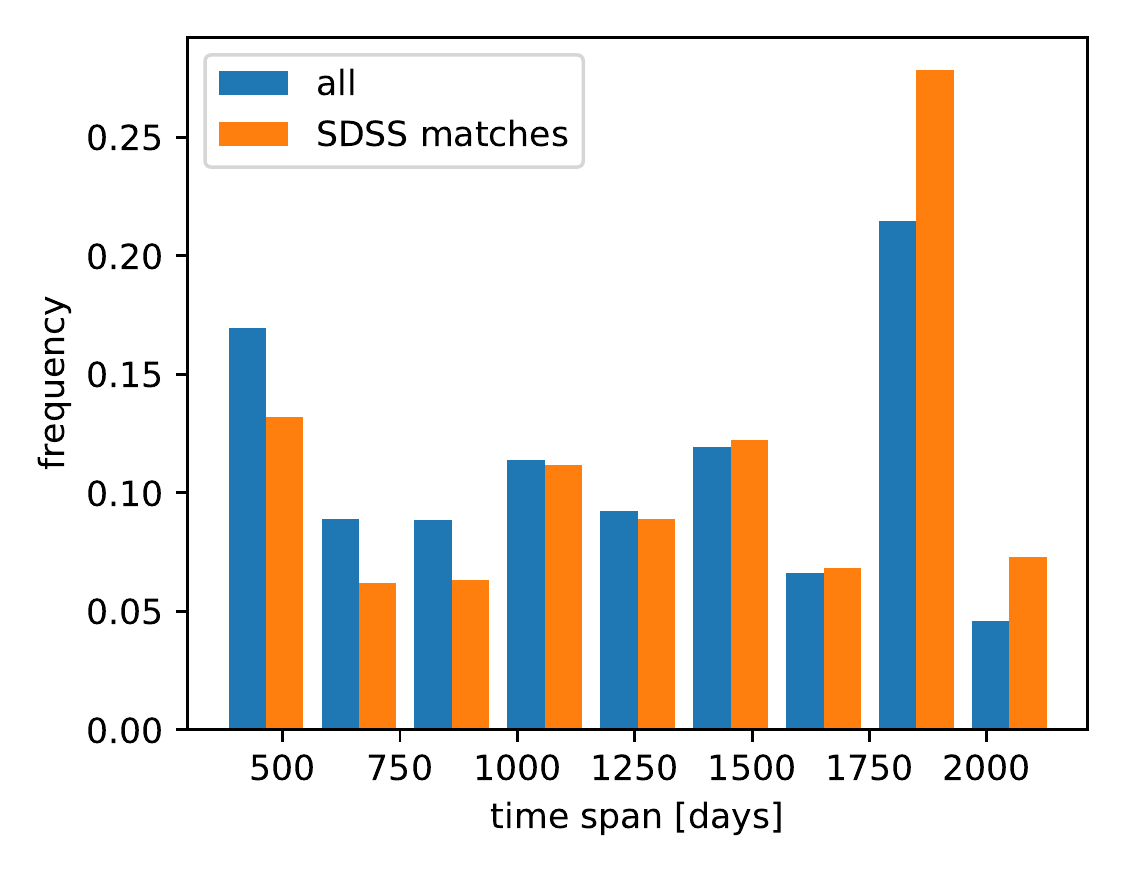}}
    \resizebox{\hsize}{!}{\includegraphics{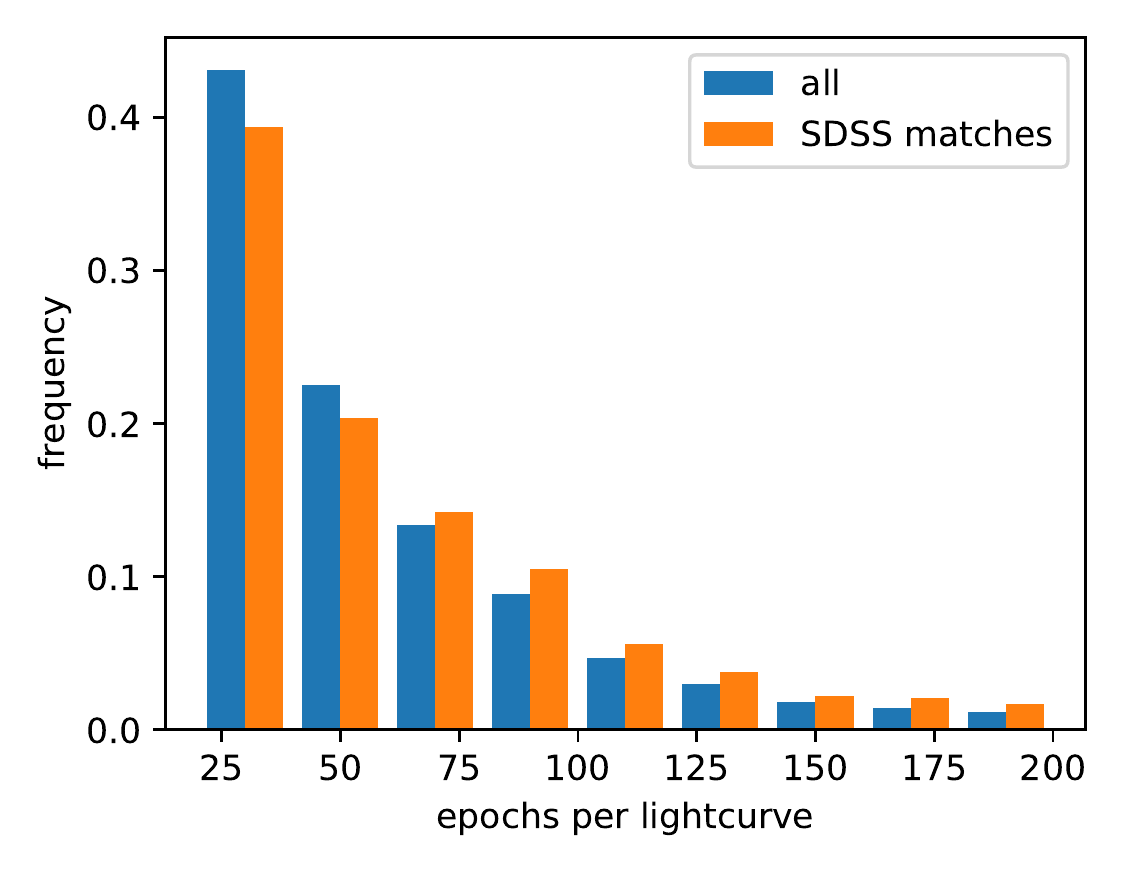}}
	\caption{Histograms of all PTF sources and sources matched to spectroscopic classifications in SDSS. The differences in time span and number of epochs show two ways in which the SDSS matched data is not representative of the entire PTF sample.} 
    \label{fig:epochs_per_lc} 
\end{figure}

\begin{figure}
	\centering
    \resizebox{\hsize}{!}{\includegraphics{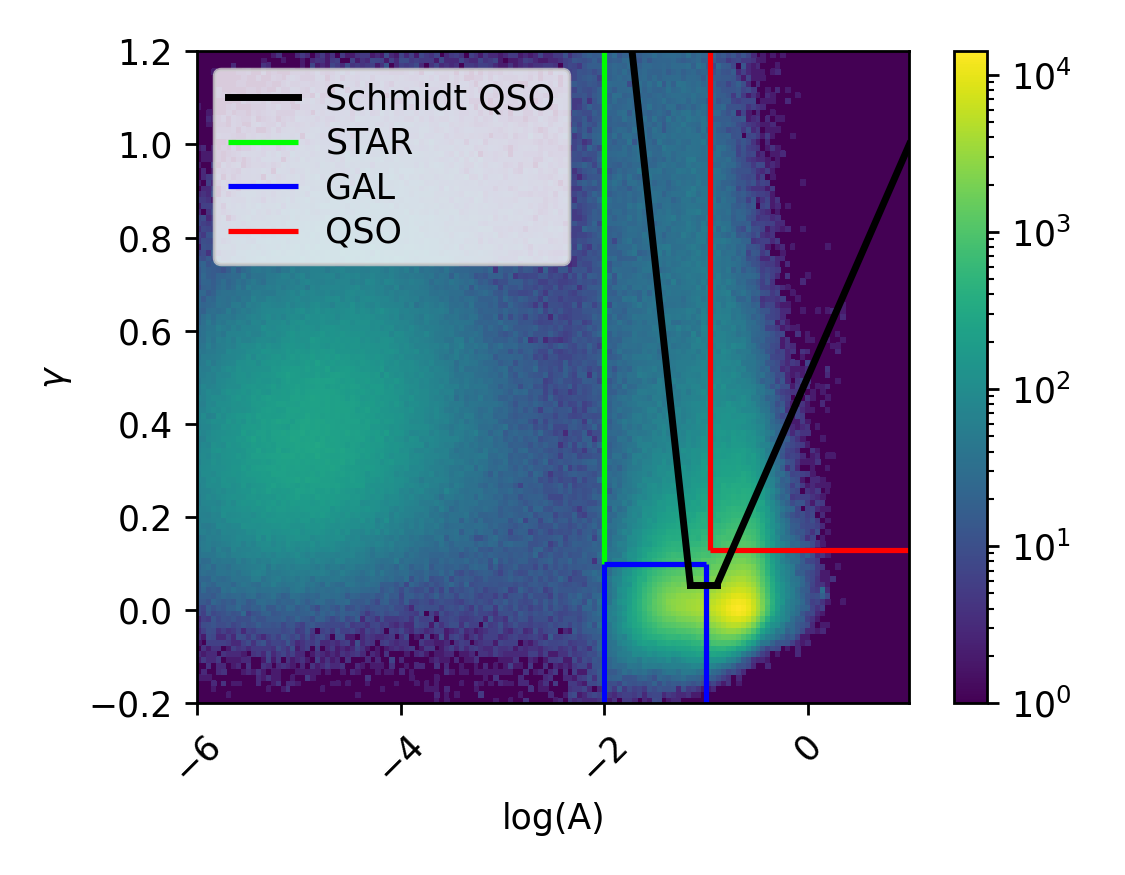}}
    \resizebox{\hsize}{!}{\includegraphics{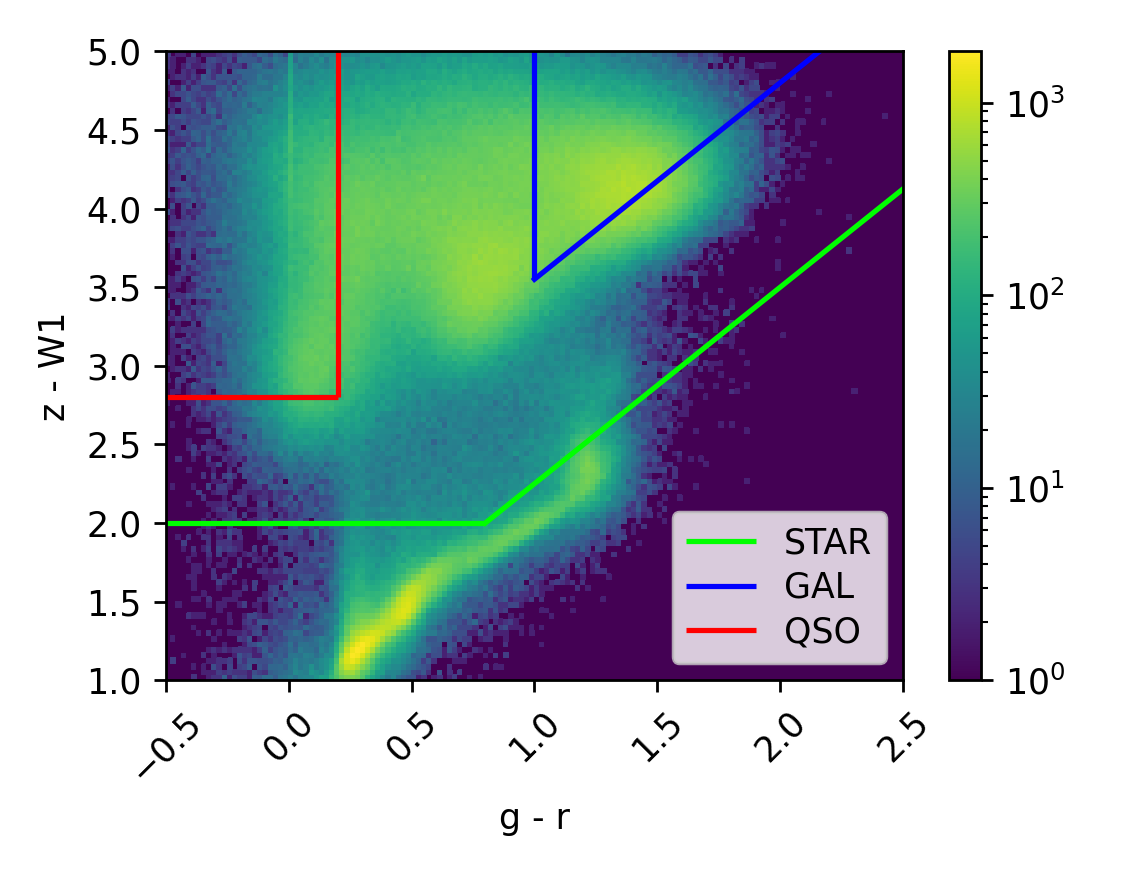}}
    
    \resizebox{\hsize}{!}{\includegraphics{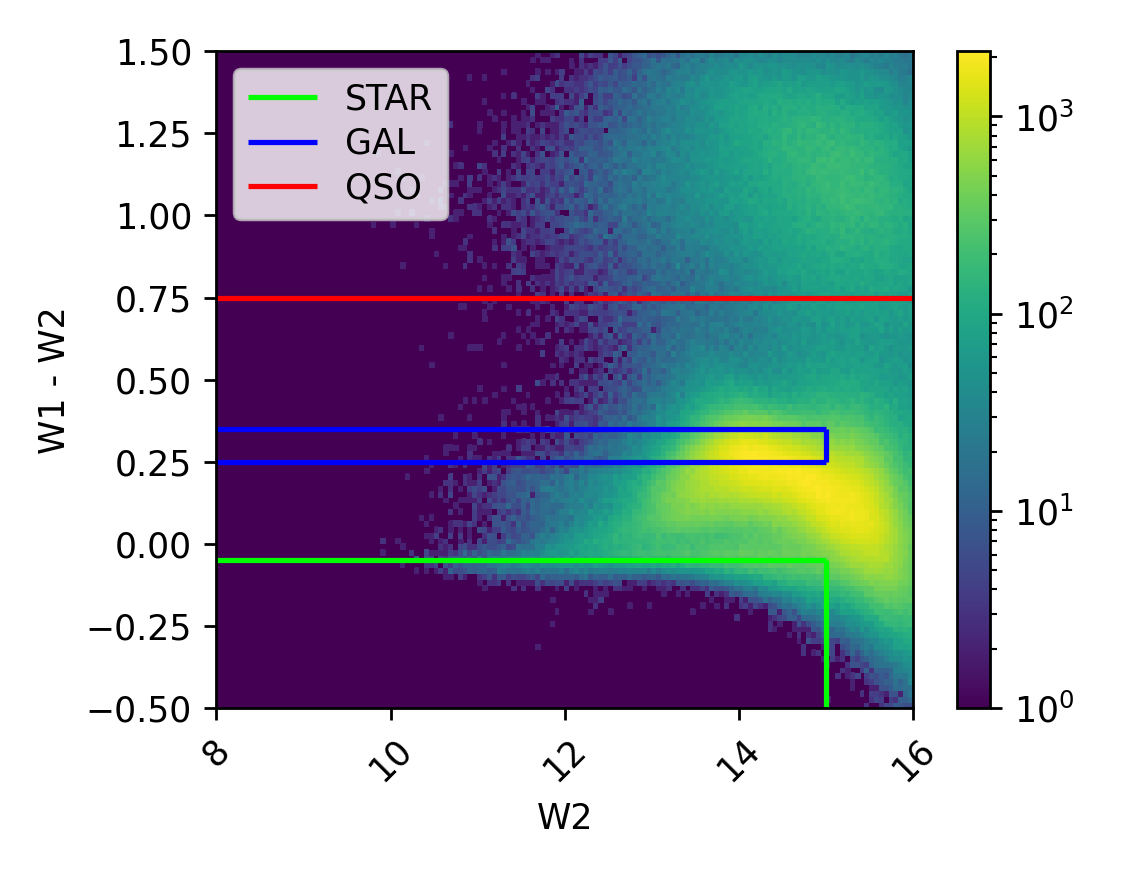}}
	\caption{Heatmaps of all data with spectroscopic classifications in SDSS. Comparing this with Fig. \ref{fig:AllData}, we see SDSS has spectroscopic data focused on specific parts of the parameter space. Strict criteria for pure selection of stars (green), QSOs (red) and galaxies (blue) are overplotted. These are listed in Table \ref{tab:thresholds}. The fit parameter panel also includes a black line showing the \citet{schmidt} QSO selection.} 
    \label{fig:heatmaps_SDSS} 
\end{figure}

\begin{figure}[H]
	\centering
    \resizebox{\hsize}{!}{\includegraphics{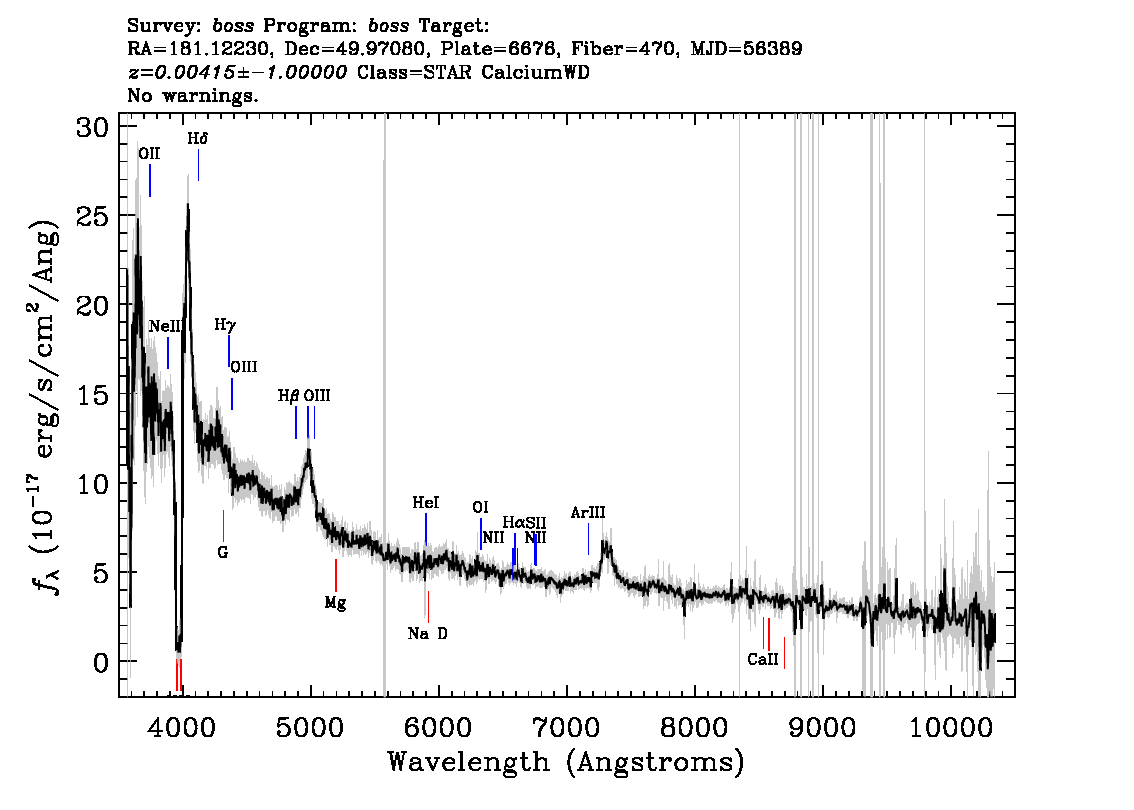}}
	\caption{Spectrum of SDSS J120429.34+495814.4 from the BOSS spectrograph. The source is classified as a star in SDSS-IV DR17, but it has the variability and colour parameters of a QSO candidate. The broad emission lines support the latter classification. This is one of 32 objects, selected as a QSO candidate and whose SDSS spectra were classified as stars, of which 20 -- 50~\% are QSOs based on visual inspection of the spectra.  (SDSS-IV DR17, CC-BY license, \url{skyserver.sdss.org/dr17/VisualTools/explore/summary?objId=1237658613058109587})} 
    \label{fig:QSO_spectrum}
\end{figure}

2.5~\% of the fitted PTF sources are matched to spectroscopic classifications in SDSS. This subset has different parameter distributions from the full PTF sample. On average, the time spans are longer and the number of epochs is higher, as shown in Fig. \ref{fig:epochs_per_lc}. The distributions of variability parameters $A$ and $\gamma$, and colours in the optical and mid-IR are shown in Fig. \ref{fig:heatmaps_SDSS} for SDSS matched data for comparison with the full sample in Fig. \ref{fig:AllData}. The differences are discussed in Sect. \ref{sec:specbias}. Fig. \ref{fig:heatmaps_SDSS} also illustrates the selection criteria for stars, galaxies and QSOs in Table \ref{tab:thresholds} and the criteria of \citet{schmidt}:
\begin{align}
    \gamma&>0.055 \label{eq:schmidt1}\\ 
    \gamma&>0.5\log_{10}A + 0.5\\
    \gamma&>-2\log_{10}A -2.25.\label{eq:schmidt3}
\end{align}
These criteria can point to potential misclassifications in SDSS, as discussed in Sect. \ref{sec:ambi} and an example of this is SDSS J120429.34+495814.4. This object is registered as a star in SDSS, but the selection criteria of this paper point to it being a QSO. The spectrum in Fig. \ref{fig:QSO_spectrum} has the broad emission lines characteristic of a QSO.

\newpage
\section{Variability selected objects} \label{app:GAL}

\begin{figure}[H]
	\centering
    \resizebox{\hsize}{!}{\includegraphics{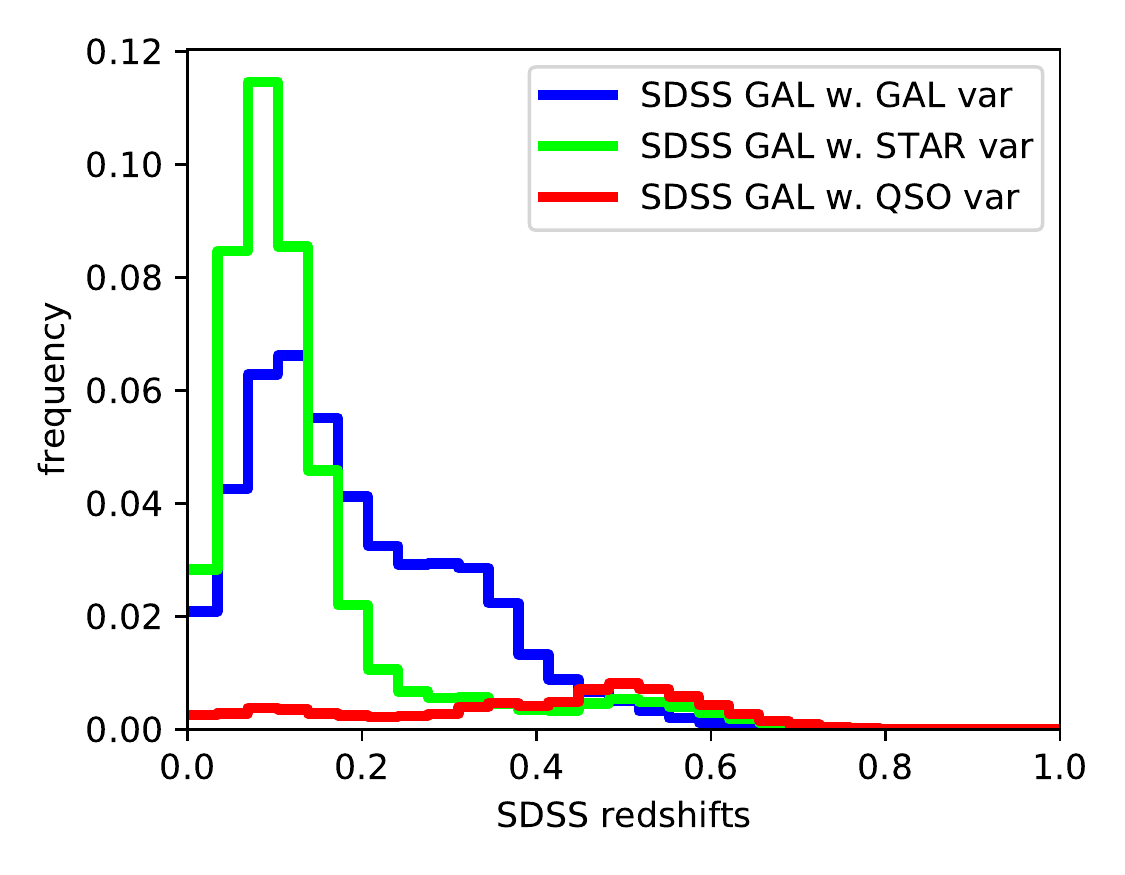}}
    \resizebox{\hsize}{!}{\includegraphics{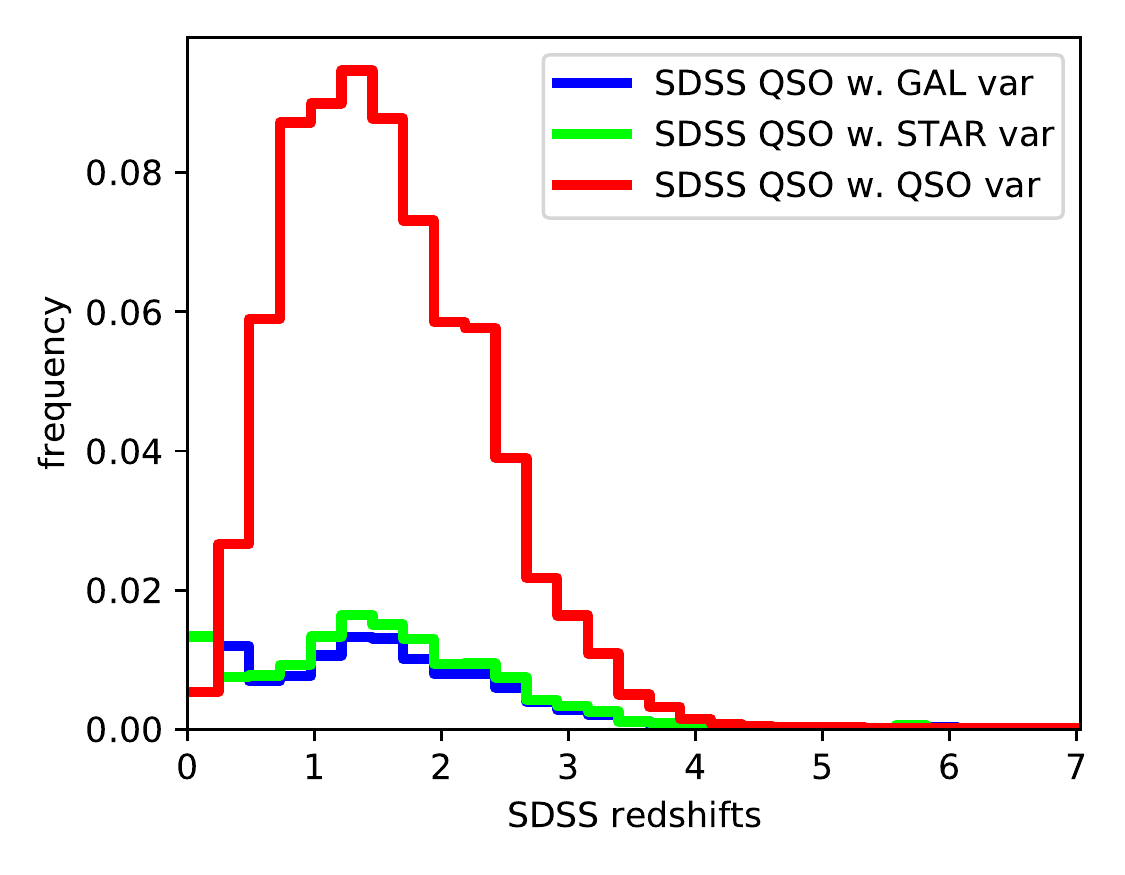}}
	\caption{Redshift distributions of galaxies (top) and QSOs (bottom) that appear as candidates of different classes based on variability criteria. The galaxies with QSO-like, high variability are typically found at high redshifts of $\sim0.5$, and those with star-like, low variability have lower redshifts than other galaxies. SDSS QSO are dominated by QSO-like variability except at $z<0.3$. The bins of each diagram are normalised with respect to the total samples of spectroscopic SDSS galaxies and QSOs, respectively.} 
    \label{fig:GAL_redshifts} 
\end{figure}

The photometric selections are performed based on either the variability criteria, the colour criteria or both of Table \ref{tab:thresholds}. Some spectroscopically confirmed objects have conflicting photometric parameters, as discussed in \ref{sec:ambi}. In Fig. \ref{fig:GAL_redshifts} the redshifts for spectroscopic galaxies are typically higher for variability selected galaxy candidates than star candidates. QSOs have similar variability except at $z<0.3$ where more appear like stars or galaxies in $A$ and $\gamma$.

\begin{figure}[H]
	\centering
     \resizebox{\hsize}{!}{\includegraphics{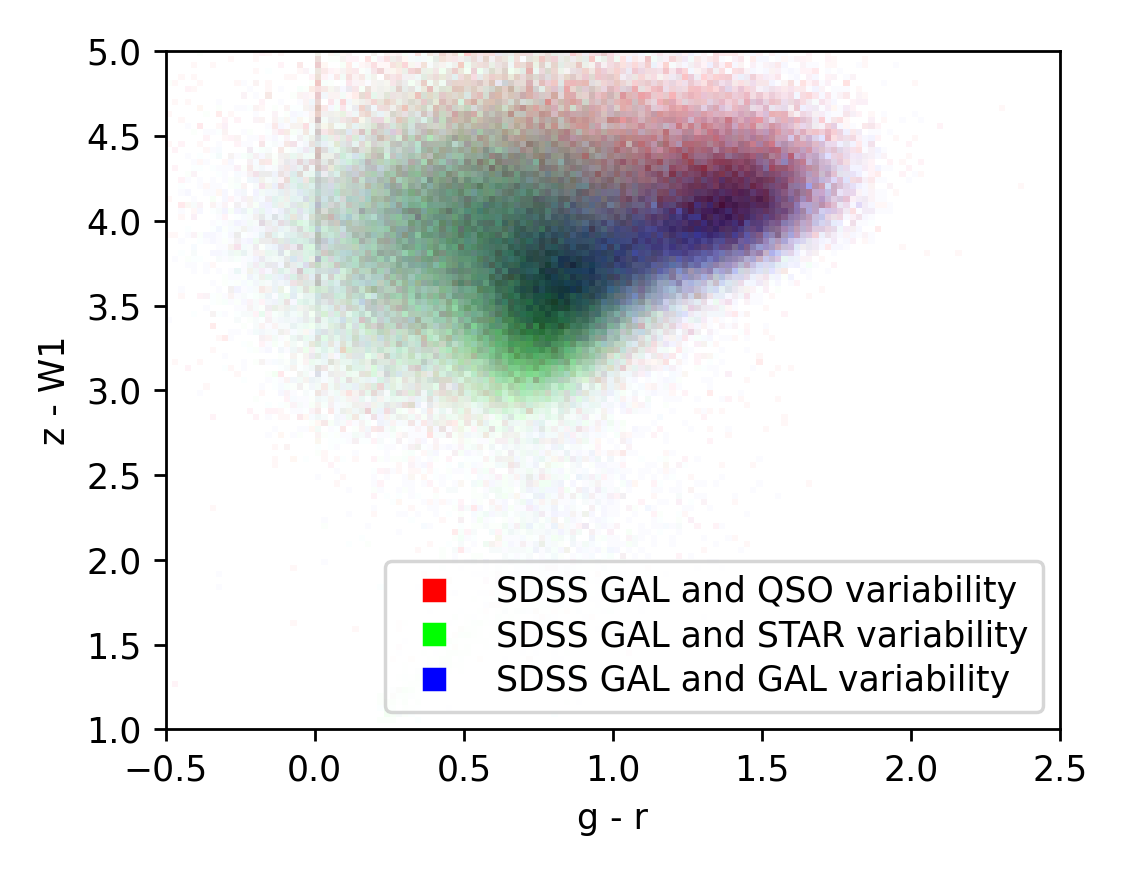}}
    \resizebox{\hsize}{!}{\includegraphics{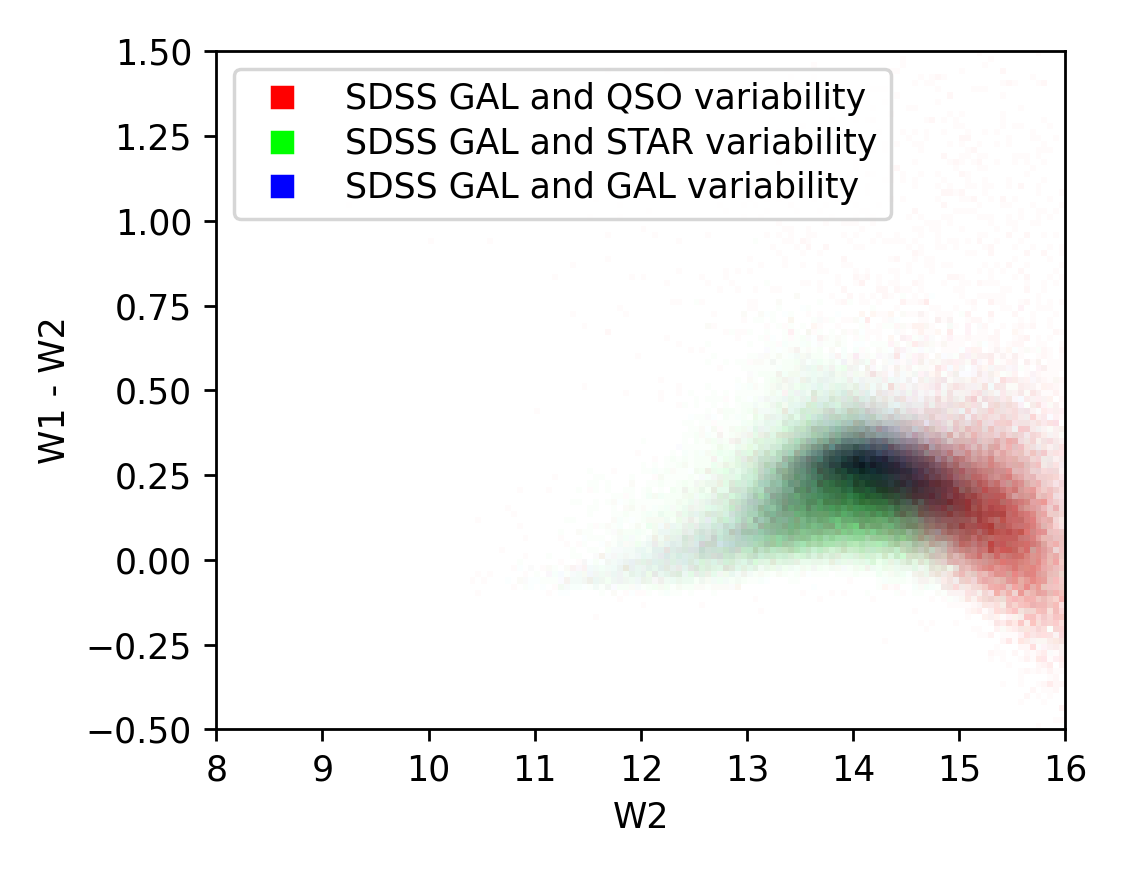}}

	\caption{Galaxies with variability parameters typical of different types of sources. We see that the more variable galaxies (with QSO variability) have high $W2$ and $z-W1$.} 
    \label{fig:GAL_var} 
\end{figure}

\begin{figure}[H]
	\centering
     \resizebox{\hsize}{!}{\includegraphics{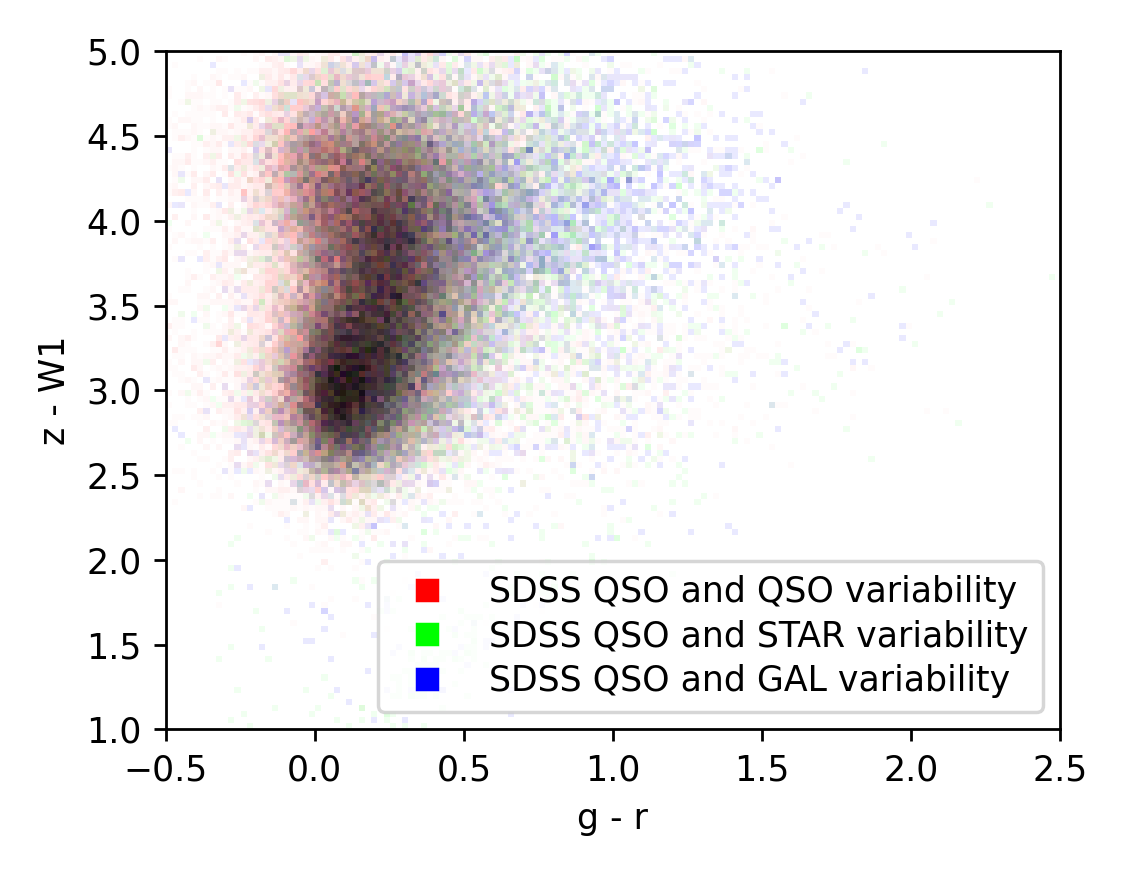}}
    \resizebox{\hsize}{!}{\includegraphics{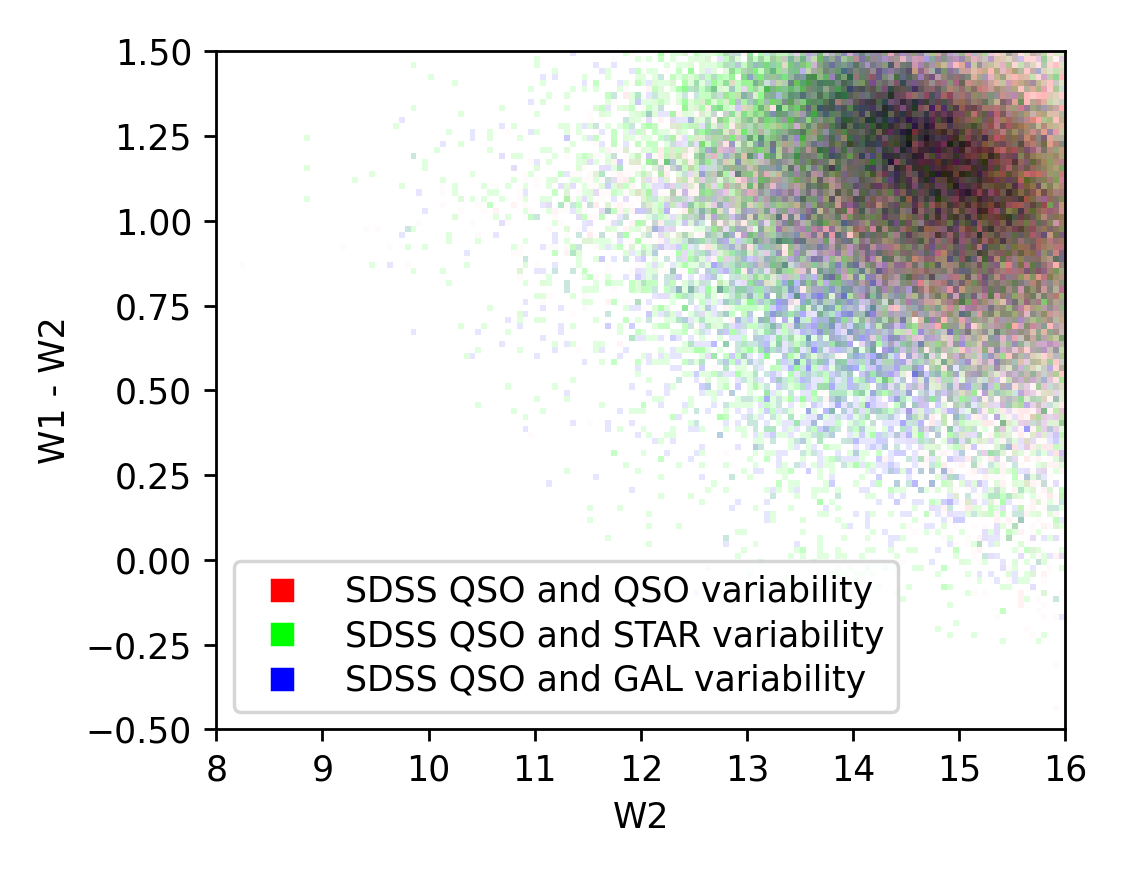}}
    
	\caption{QSOs with different variability parameters are still mostly found in the same regions of the colour diagrams.} 
    \label{fig:QSO_var} 
\end{figure}

\begin{figure}[H]
	\centering
    \resizebox{\hsize}{!}{\includegraphics{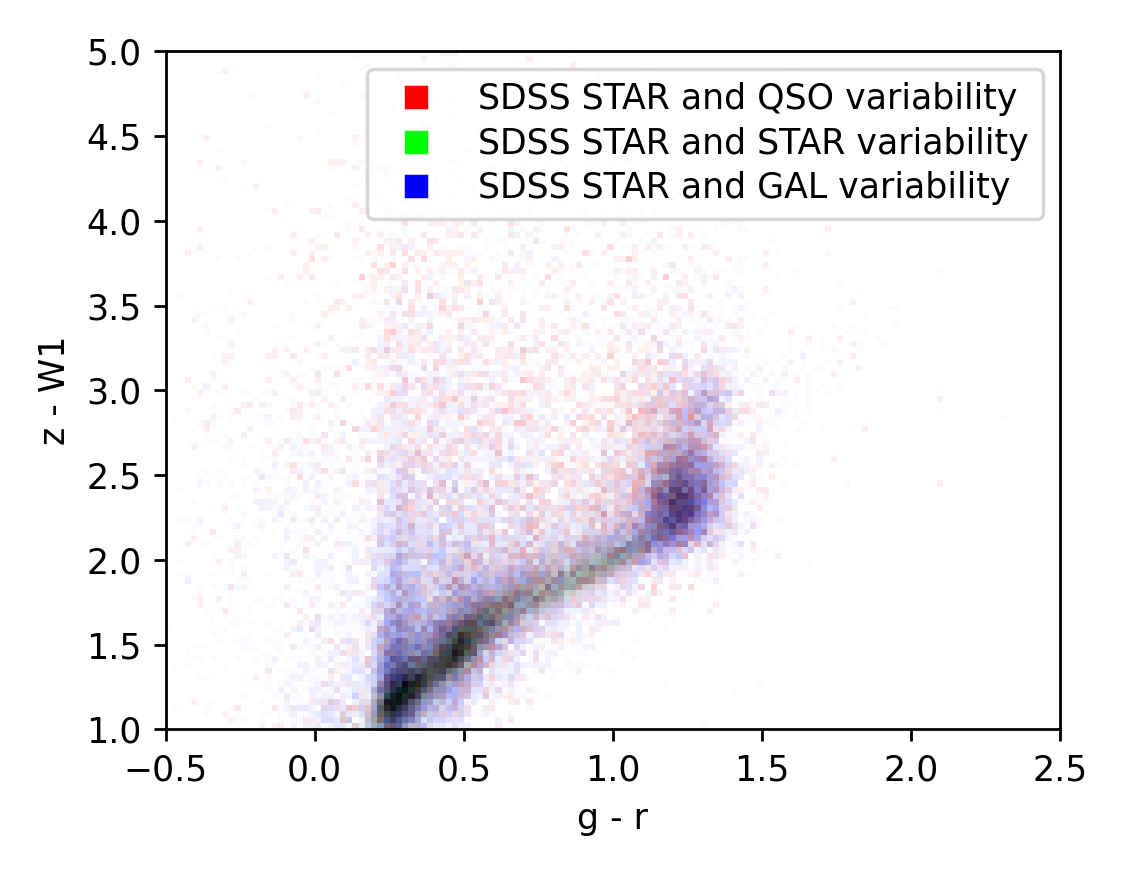}}
    \resizebox{\hsize}{!}{\includegraphics{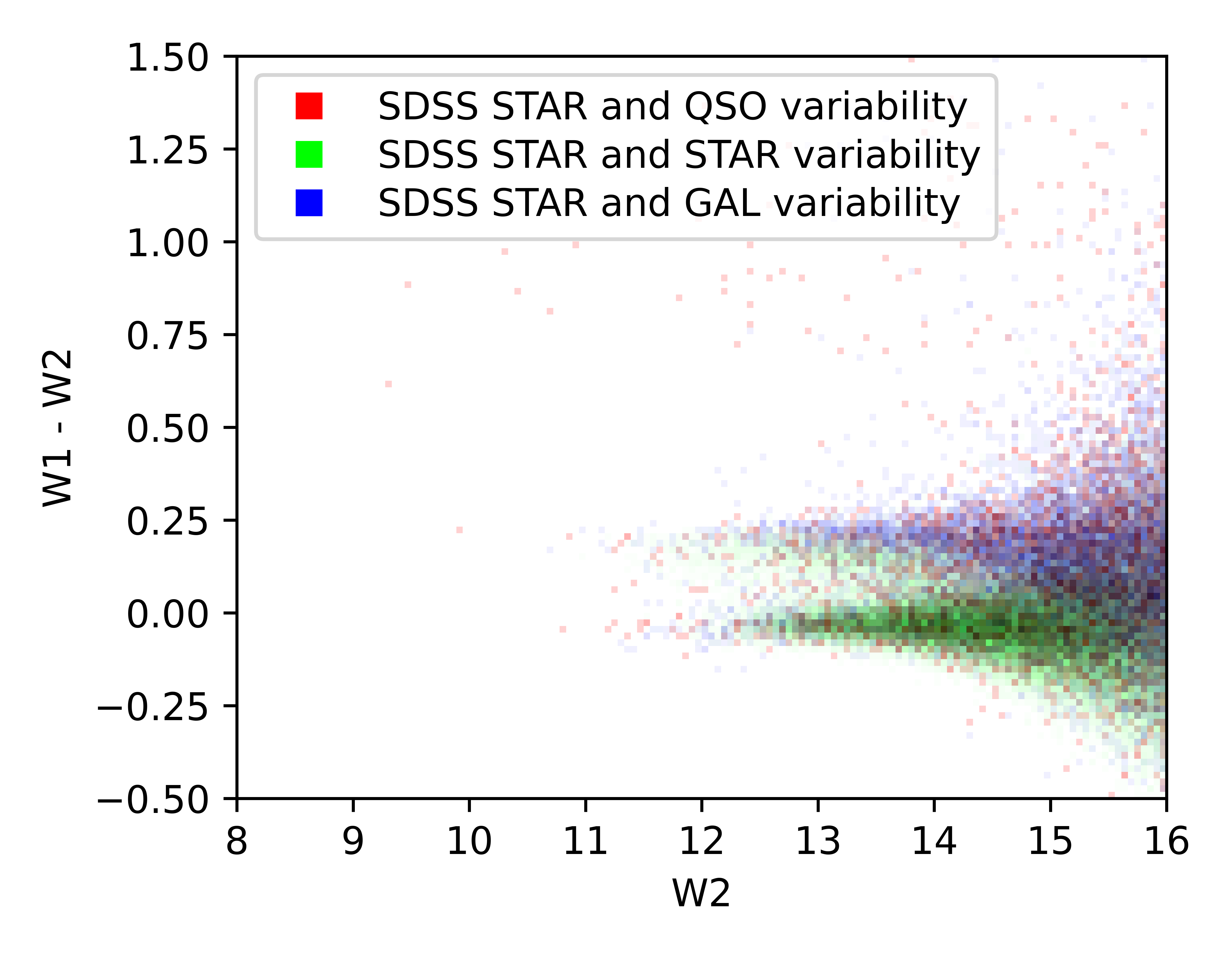}}
    
	\caption{Stars are in the stellar locus of the upper panel, but with more spread for those with variability that is more typical of QSOs and galaxies. Galaxy-like variability is also found at higher $W1-W2$ than for most stars. Based on inspection of SDSS imaging, the 73 SDSS stars with galaxy-like colours mostly look like galaxies or a star-galaxy chance alignment. SDSS stars with $0.2<W1-W2>0.3$ and $W2<13$ are also often not isolated. We do not see this for random subsets of all of the 12\,470 SDSS stars with galaxy-like variability.  } 
    \label{fig:STAR_var} 
\end{figure}

Figs. \ref{fig:GAL_var}--\ref{fig:STAR_var} show colour diagrams for all combinations of spectroscopic and variability selected classes. This illustrates how the variability selection criteria are picking objects with different colour distributions, even when the objects are spectroscopically confirmed to belong to the same class. This is discussed further in Sects. \ref{sec:ambi} and \ref{sec:specbias}.

\newpage
\section{SIMBAD statistics} \label{app:SIMBAD}

\begin{figure}[H]
	\centering
    \resizebox{\hsize}{!}{\includegraphics{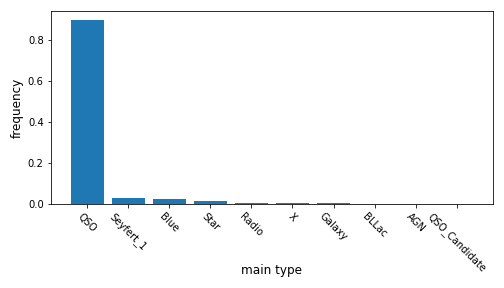}}
    \resizebox{\hsize}{!}{\includegraphics{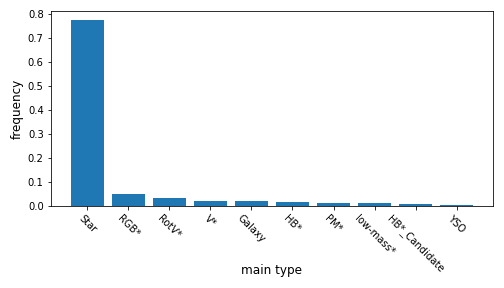}}
    \resizebox{\hsize}{!}{\includegraphics{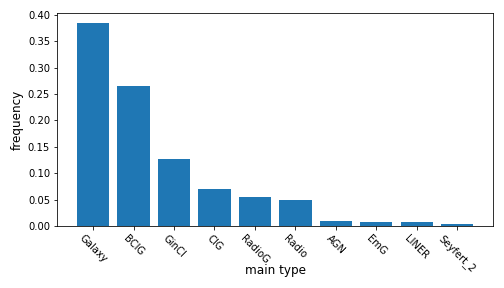}}
    
	\caption{SIMBAD "main type" statistics for candidates by colour and variability. This shows the subtypes and contaminating sources for QSO (upper panel), star (middle) and galaxy (lower) candidates.} 
    \label{fig:SIMBAD_colvarclasses} 
\end{figure}

Candidate stars, QSOs and galaxies are selected based on the criteria in Table \ref{tab:thresholds}. To understand the physical nature of the objects, we looked up statistics of their "main type" in SIMBAD. Most photometrically selected objects were registered with the an expected label in SIMBAD, as shown in Fig. \ref{fig:SIMBAD_colvarclasses}. 

\section{\textit{Gaia} proper motions} \label{app:PM}

\begin{figure}[H]
	\centering
    \resizebox{\hsize}{!}{\includegraphics{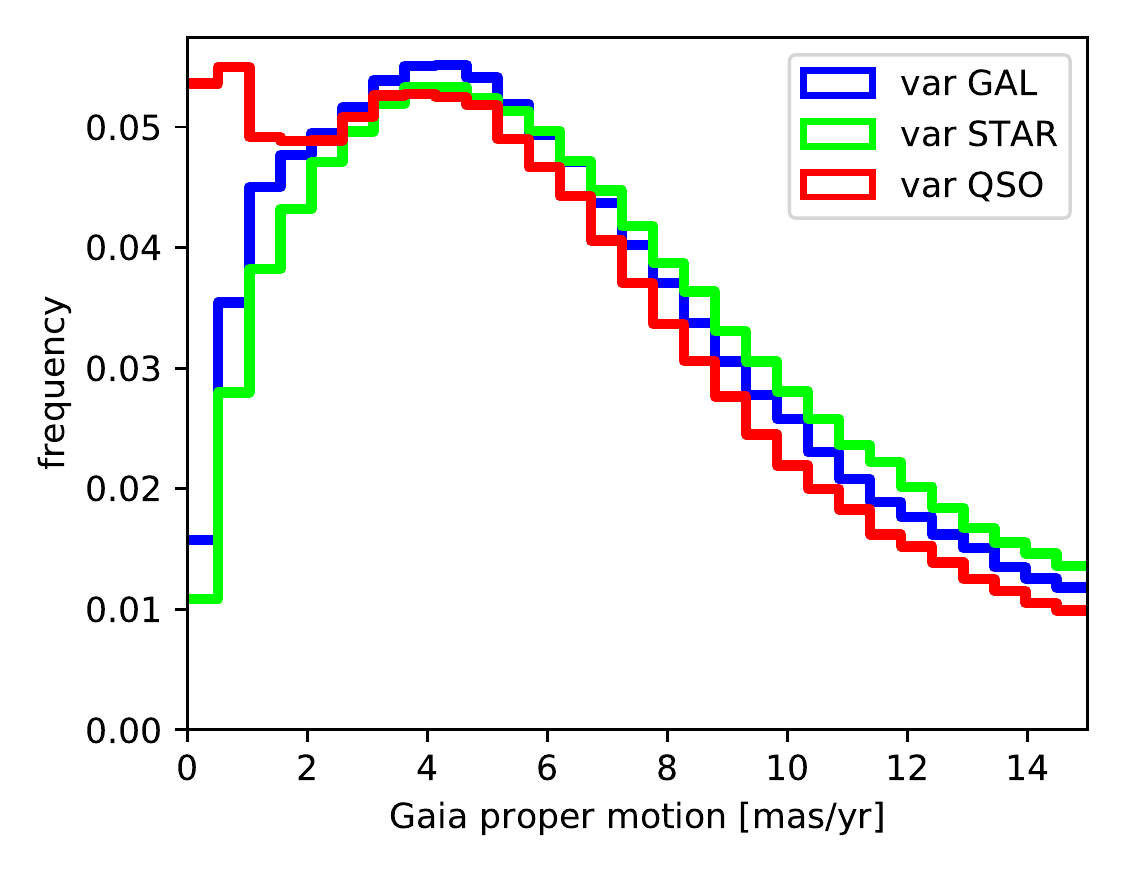}}
    \resizebox{\hsize}{!}{\includegraphics{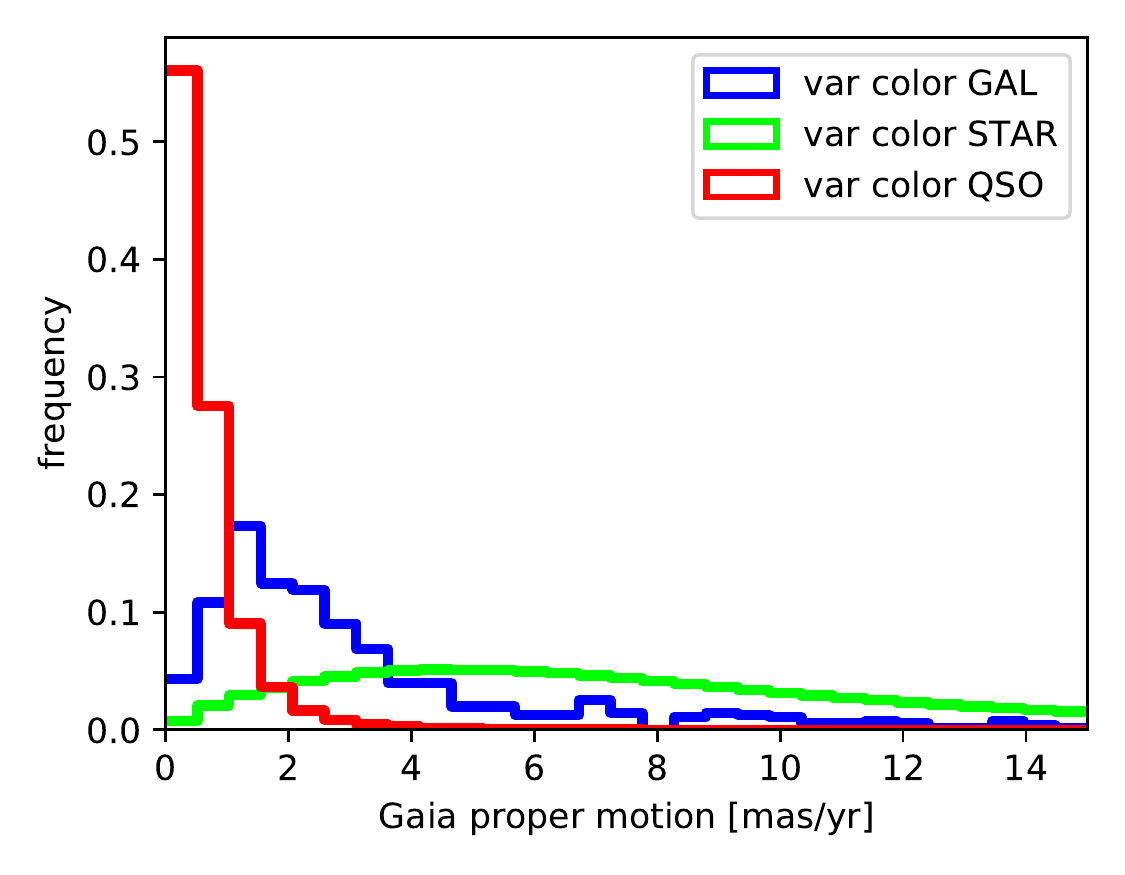}}
    \resizebox{\hsize}{!}{\includegraphics{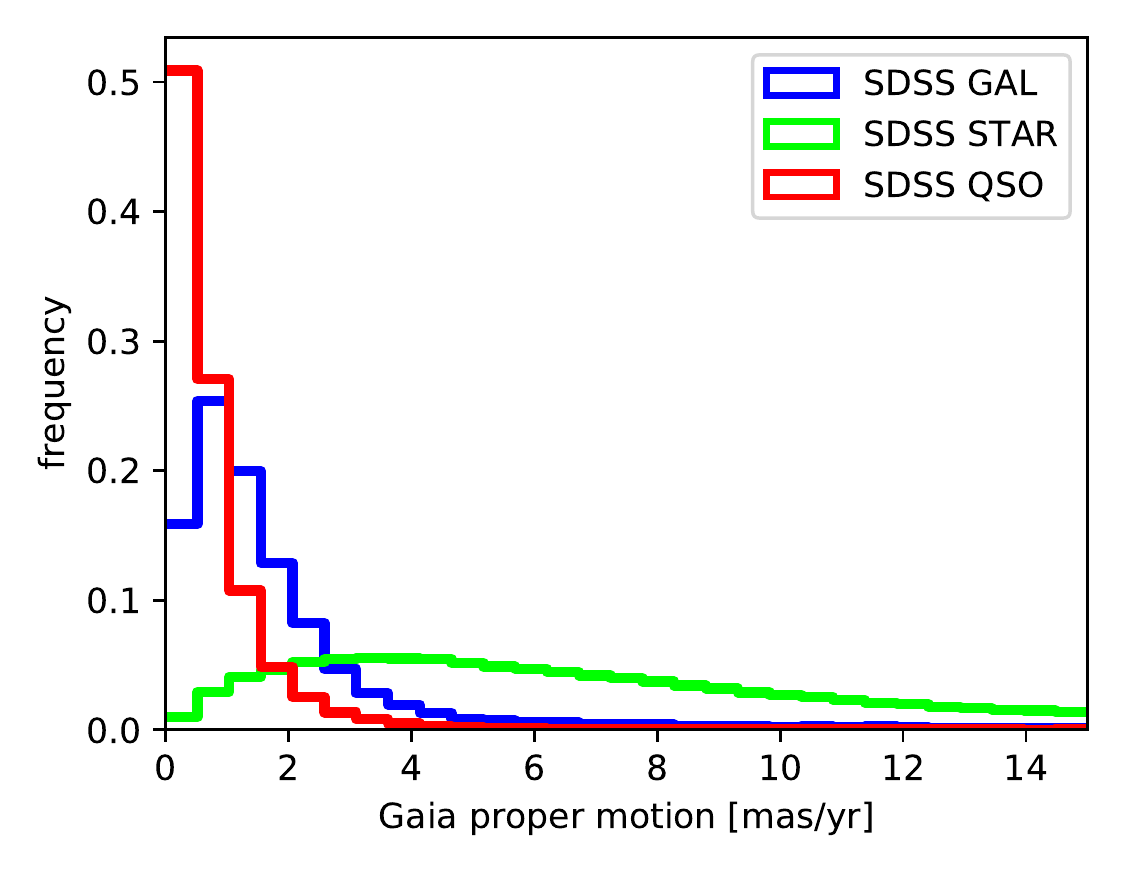}}
	\caption{\textit{Gaia} proper motions for sources selected by variability (top), colour and variability (middle) and spectroscopy (bottom). 
 The histograms are normalised individually. All variability-selected sources approximately follow the distribution of SDSS stars, but when we also select by colour, the results are generally close to those of the spectroscopically confirmed objects. 
 A high-purity and high-completeness cut at $\lesssim 2$ mas/y, justified by the bottom panel, would yield $\approx 50\%$ purity in a sample of quasar candidates selected only through variability.}
    \label{fig:GaiaPM} 
\end{figure}

We cross match with \textit{Gaia} DR3 \citep{Gaia,GaiaDR3} proper motions using the CDS X-Match Service \citep{Xmatch}. This is done for all sources within one arcsecond of the PTF coordinates. For those selected through variability, we limit the search to 1 million random sources of each class. In Fig. \ref{fig:GaiaPM} we show histograms of the sources with measured proper motions. As expected, these are generally lower for QSOs and higher for stars. The differences are smaller for variability-selected objects and they follow the distribution of spectroscopically confirmed stars almost perfectly, indicating that most of the objects with non-zero proper motions are stars. However, the objects with QSO-like variability have a higher spectroscopic purity (59.3~\%) in the SDSS. Sources selected by both colour and variability have distributions closer to those of the spectroscopically confirmed classes, especially for QSOs. Note that the sample sizes are also different, as not all sources have a match in \textit{Gaia} and multiple \textit{Gaia} sources can be detected per PTF source. For variability-selected objects, the \textit{Gaia} sets are
60.6~\%, 92.0~\% and 72.0~\% 
the sizes of the QSO, star and galaxy sets, respectively. For sources selected in variability and colour, the numbers are 
84.7~\%, 99.6~\% and 0.5~\%. 
For spectroscopically confirmed sources, they are 
82.3~\%, 95.9~\% and 3.7~\%. 
\textit{Gaia} has proper motions for almost all stars and very few galaxies, as its design is optimised towards detecting stars and the limit for proper-motion measurements is not as faint as the PTF or Pan-STARRS source-detection depth. As is shown by the comparison of SDSS and \textit{Gaia} purity of the variability-selected QSOs, atypical proper motions may indicate misclassifications and mismatches between surveys, or interesting sub-classes or flaring objects. The sum of the residuals between histograms of variability-selected QSOs and variability-selected stars (-0.01 for stars to calibrate the curves at high proper motions), when the total area under the curve is normalised to 1, is 0.29. This gives a lower bound on the QSO purity, in addition to the higher spectroscopic purity, as it does not take into account differences in how many objects are matched, chance alignments of stars etc. (even SDSS QSOs usually have a proper motion measured nearby in \textit{Gaia}).

\end{appendix}

\end{document}